\documentclass[prd,a4paper,aps,showpacs,twocolumn,floats,nofootinbib]{revtex4}
\usepackage{epsfig}
\usepackage{graphicx,amsmath,amsfonts,amssymb,slashed,upgreek,color}
\usepackage{enumerate}

\newcommand{\bk}{{\mathbf k}}

\newcommand{\bq}{{\mathbf q}}

\newcommand{\bx}{{\mathbf x}}

\newcommand{\al}{\alpha}

\newcommand{\de}{\delta}
\newcommand{\De}{\Delta}
\newcommand{\ep}{\epsilon}
\newcommand{\ga}{\gamma}

\newcommand{\la}{\lambda}
\newcommand{\Om}{\Omega}

\newcommand{\be}{\begin{equation}}
\newcommand{\ee}{\end{equation}}

\newcommand{\lsim}{\stackrel{<}{\sim}}
\newcommand{\bea}{\begin{eqnarray}}
\newcommand{\eea}{\end{eqnarray}}
\newcommand{\bean}{\begin{eqnarray*}}
\newcommand{\eean}{\end{eqnarray*}}

\newcommand{\HH}{{\cal H}}

\newcommand{\vev}[1]{\mbox{$\langle #1 \rangle $}}

\newcommand{\eb}{{\rm em}}

\newcommand{\omem}{{\Omega^{-}_{\rm em}}}
\newcommand{\ombp}{{\Omega^{+}_{\rm B}}}
\newcommand{\omb}{{\Omega_{\rm B}}}
\newcommand{\omp}{{\Omega_\Pi}}
\newcommand{\ompm}{{\Omega^{-}_\Pi}}
\newcommand{\ompp}{{\Omega^{+}_\Pi}}

\newcommand{\zetainf}{{\zeta_{\rm inf}}}

\begin{document}

\title{Magnetic fields from inflation: The CMB temperature anisotropies}

\author{Camille Bonvin$^{1,2}$, Chiara Caprini$^{3}$ and Ruth Durrer$^{4}$
\\  }
\affiliation{$^{1}$ Kavli Institute for Cosmology Cambridge and Institute of Astronomy,
Madingley Road, Cambridge CB3 OHA, United Kingdom\\
${}^{2}$ DAMTP, Centre for Mathematical Sciences, Wilberforce Road, Cambridge CB3 OWA, United Kingdom\\
${}^{3}$CEA, IPhT and CNRS, URA 2306, F-91191 Gif-sur-Yvette, France \\
${}^{4}$D\'epartement de Physique Th\'eorique and Center for Astroparticle Physics, Universit\'e de 
Gen\`eve, 24 quai Ernest 
Ansermet,~CH--1211 Gen\`eve 4, Switzerland}

\date{\today}

\begin{abstract}
The cosmic microwave background (CMB) temperature anisotropies from primordial magnetic fields are studied. In addition to the known passive and compensated modes we discuss an inflationary magnetic mode in the curvature perturbation, present when magnetic fields are generated during inflation. This mode is absent if the generation mechanism is causal, e.g. a phase transition. We compute and discuss the effect of this mode on the observed CMB anisotropy spectrum, in comparison with the passive and compensated ones. We find that it dominates the CMB anisotropy, and consequently leads to stronger constraints on the amplitude $B_{1\,{\rm Mpc}}$ and spectral index $n_B$ of the magnetic field than what usually found in CMB analyses from the compensated mode. This happens in particular for spectral indexes $n_B>-3$: the inflationary magnetic mode is always scale invariant, therefore through this mode even a magnetic field with a spectrum which is not scale invariant can leave a detectable signal in the CMB at large scales. 
\end{abstract}


\maketitle

\section{Introduction}
\label{sec:intro}

The effects of a stochastic primordial magnetic field on the CMB are many: spectral distortions of the monopole~\cite{Puy:1998sv,Jedamzik:1999bm}, the generation of scalar, vector and tensor perturbations in the metric affecting both the temperature and the polarization spectra, see e.g. \cite{Durrer:1999bk,Caprini:2003vc,Paoletti:2012bb,Paoletti:2010rx,finelli,Shaw:2010ea,shaw,Lewis:2004ef,Kunze:2010ys,Giovannini:2009zq,Giovannini:2007qn,Yamazaki:2011eu,Yamazaki:2012pg,Durrer:2006pc}, the production of non-Gaussian signatures leading to a non-zero temperature bispectrum and trispectrum, see e.g. \cite{Seshadri:2009sy,Caprini:2009vk,Shiraishi:2012rm,Shiraishi:2010yk,Cai:2010uw}, Faraday rotation of CMB polarization, see e.g. \cite{Kosowsky:2004zh,Yadav:2012uz}. The most recent observational constraint from CMB temperature anisotropies on the amplitude and the spectral index of a primordial magnetic field has been established using Planck data \cite{Ade:2013zuv}:  
\be
B_{1\,{\rm Mpc}}<3.4~{\rm  nG}~~~{\rm with}~~~n_B<0\,.
\label{Planckconstr}
\ee 
This limit has been derived including the magnetic contribution to scalar and vector perturbations, and performing a Markov chain Monte Carlo (MCMC) analysis of the temperature angular power spectrum. 

In general, the magnetic field is modeled as a Gaussian random field, statistically homogeneous and isotropic, with a power law spectrum
\bea
\vev{B_i(\bk,\eta)B^*_j(\bq,\eta)}&=&(2\pi)^3\de^3(\bk-\bq) P_{ij} P_B(k,\eta) ~~~~~\label{mpowerspectrum}\\
P_B(k,\eta)&=& \left\{ \begin{array}{ll} 
A_B(\eta)\,k^{n_B}  & k\leq k_D(\eta) \\
 0 &  k>  k_D(\eta)\, , \end{array} \right.
  \label{PB}
\eea
where $P_{ij}=\de_{ij}-\hat{k}_i\hat{k}_j$ and $k_D(\eta)$ is a cutoff scale due to the dissipation of magnetic energy in the cosmic plasma, which has been first calculated in Refs.~\cite{Jedamzik:1996wp,Subramanian:1997gi}. The quantity in terms of which the CMB bounds are customarily expressed is the magnetic field amplitude smoothed over a comoving scale $\lambda$, set to $1$ Mpc in the Planck analysis~\cite{Ade:2013zuv}: 
\bea
B_\la^2 &=&\frac{1}{\pi^2}\int dk\,k^2\,P_B(k,\eta_0)\,e^{-k^2\la^2} \nonumber \\
&= &\frac{A_B(\eta_0)}{2\pi^2}\frac{\Gamma[(n_B+3)/2]}{\la^{n_B+3}} \label{Bla}\,.
\eea
Here $\eta_0$ denotes the conformal time today. 

The magnetic field, its power spectrum, and the upper cutoff, $k_D$, depend on time, due not only to the expansion of the universe but also due to the interaction of the magnetic field with the cosmic plasma (for a review, see \cite{ruthreview}). The  contribution to the time evolution from interactions with the cosmic plasma, i.e. MHD cascades, small scale damping  by viscosity and so on, is usually neglected in CMB analyses (with the exception of \cite{Shaw:2010ea}). In fact, these processes operate mainly at small scales, as opposed to the large scales probed by the CMB. On the other hand, the non-trivial time evolution may affect CMB analyses in the large $\ell$ range, as probed by Planck, or whenever one accounts for helical magnetic fields undergoing large scale inverse cascades (see \cite{ruthreview} and references therein). 

The magnetic field model presented above has been adopted in CMB analyses since it has the advantage to be simple and general. Only two parameters enter in the magnetic field description: once the choice of $\la$ has been made, these can be cast in the couple ($B_\la\,,\,n_B$). Effectively, when relevant, the damping scale $k_D$ can be expressed in terms of these - see e.g. \cite{Caprini:2009vk}. Therefore the constraints on the couple ($B_\la\,,\,n_B$) are, at least in principle, model independent. In particular, there is no need to specify the mechanism of generation of the magnetic field, nor its generation epoch. 

The generation of a primordial magnetic field of the order of the nanogauss is severely constrained. Causal generation mechanisms give rise to blue magnetic spectra $P_B(k)\propto k^2$ peaked on very small scales, which, accordingly, are very small at typical CMB scales~\cite{Durrer:2003ja}. Generation based on the violation of conformal invariance during inflation can lead to scale invariant spectra and relevant magnetic field amplitudes, but has in general other problems (such as strong coupling, gauge symmetry breaking or the presence of ghosts, c.f. \cite{demozzi,Himmetoglu:2009qi}). 

Nonetheless, the aim of the CMB analyses carried out so far is to constrain the presence of a primordial magnetic field in a {\it model independent way}, regardless of what is easy to produce or natural to expect; therefore, in these analyses one adopts a magnetic field model which is sufficiently general. 

In practice, however, the situation is somewhat more involved. The generation mechanism of the magnetic field affects not only its spectral index, which is one of the parameters constrained by MCMC analyses of the CMB, but also the initial conditions of the Boltzmann hierarchy right after neutrino decoupling. Up to now, CMB data analyses have only taken into account the so-called {\em compensated mode} \cite{Paoletti:2010rx,finelli,shaw,Shaw:2010ea,Paoletti:2012bb,Giovannini:2008yz}: a particular set of initial conditions giving rise to an isocurvature ($\zeta=0$) mode, in which the fluid and magnetic energy densities and anisotropic stresses are compensated. This mode is one of the possible solutions of Einstein's equations with free-streaming neutrinos, and it is independent of the way the magnetic field is generated. 

However,  previous analyses have shown that there are other perturbation modes from magnetic fields that add to the compensated mode. There is first the so-called {\em passive mode}, which is an adiabatic-like mode that depends logarithmically on the time of generation of the magnetic field $\eta_*$. Its contribution to the CMB spectrum is  in general larger than the one of the compensated mode (c.f. section \ref{sec:angular} and Refs.~\cite{shaw,magcaus}). Accounting for this mode would therefore change the bound in Eq.~\eqref{Planckconstr}. Or at least, given the Planck bound for the compensated mode, it adds a constraint on the new parameter: $\eta_*$, i.e. it constrains the time of magnetic field generation. 

The purpose of this paper is to show the effect on the CMB of yet another mode, which is also adiabatic, and is present only if the magnetic field is generated during inflation. We call it the {\em inflationary magnetic mode}. The existence of this mode has been demonstrated in Refs.~\cite{maginf,barnaby,seery}, therefore this paper is  intended as a follow-up and complement of Ref.~\cite{maginf}. 

The inflationary magnetic  mode is distinctively different from the compensated and the passive modes, since the curvature is dynamically generated only during inflation, and remains constant in the radiation/matter era (just as the usual inflationary mode due to the quantum fluctuations of the inflaton field). Moreover, it does not depend directly on the magnetic field power spectrum: we will show that through this new mode, even a magnetic field which is far from scale invariance can leave a detectable imprint on the CMB at large scales. The distinction of such a mode from the usual inflationary mode comes mainly from its statistics which is non-Gaussian~\cite{barnaby}. Another possible difference with respect to the adiabatic mode from inflation is that this mode can have logarithmic corrections to scale invariance, see Eq.~(\ref{zetainf}).

In this paper we show that this mode, if it is present, generically dominates the contributions to the CMB temperature perturbations due to the magnetic field. Therefore, it should be taken into account when constraining primordial magnetism with CMB data. As we shall see, this implies, however, to insert the magnetic field generation time, or the redshift of reheating, as an extra parameter in the magnetic field model, and diversify the CMB constraints depending on the mechanism of generation of the magnetic field. On the other hand, if magnetic fields are generated during inflation and this mode is present, it provides in principle a new way to determine the energy scale of inflation, which (in the simple approximation used here) directly determines the reheating temperature.  

The remainder of the paper is structured as follows: in~Section \ref{sec:inf}, we revisit  the model of inflationary magnetogenesis which we adopt in the following. We basically summarise the results of Ref.~\cite{maginf}, showing the effect of an inflationary magnetic field on the comoving curvature perturbation $\zeta$ at super-horizon scales. In Section~\ref{sec:psirec}, we compute the metric perturbations analytically at super-horizon scales until recombination. In Section~\ref{sec:SW}, we evaluate analytically the Sachs Wolfe effect, i.e. the temperature anisotropy at large scale and at recombination time, from the compensated, passive and the new inflationary magnetic mode. In Section \ref{sec:angular} we present the different CMB spectra evaluated both analytically and using the CAMB code~\cite{shaw}, and we compare the contributions of the different modes. In Section~\ref{sec:discussion} we discuss our results and we conclude in Section~\ref{sec:con}. 

{\bf Notation:} Throughout this paper we use conformal time $\eta$, comoving space coordinates ${\bf x}$ and wave vectors ${\bf k}$ with the spatially flat metric $ds^2=a^2(\eta)(-d\eta^2+ \de_{ij}dx^i dx^j)$; greek letters denote 4d spacetime indices while latin letters denote 3d spatial indices and spatial vectors are denoted in bold face. For the metric and scalar field perturbations we follow the conventions of~ \cite{mukhanov}.  An overdot denotes derivatives with respect to conformal time $\eta$, and a prime with respect to the variable $x=|k\eta|$. We define the Planck mass by  $m_P = (\sqrt{8\pi G})^{-1}$.

\section{Inflationary generation mechanism}
\label{sec:inf}

In this section we show that a primordial magnetic field which is generated during inflation leads to a new mode in the initial conditions for the evolution of metric perturbations after inflation. Specific examples of inflationary generation mechanism are worked out, e.g., in 
Refs.~ \cite{turner,ratra,Giovannini:2000dj,Dimopoulos:2001wx,bamba1,Anber:2006xt,bamba2,yokoyama,subramanian,Durrer:2010mq}.  

The existence of this mode is due to the fact that the magnetic energy momentum tensor gravitates and perturbs the metric. The mode is therefore present for any model of magnetic field generation from inflation, and choosing a specific model only changes the details but not the substance of this analysis, which can be easily generalized to other generation mechanisms. We consider the simplest existing model for magnetic field generation from inflation~\cite{ratra}, even though it has been shown to suffer from a strong coupling problem~\cite{demozzi} (a possible way to avoid the strong coupling problem is proposed e.g. in~\cite{Ferreira:2013sqa}). As shown in the following, the specific form of the coupling determines the power spectra of the magnetic energy density and of the anisotropic stress. We keep their amplitude and spectral index as free parameters throughout the analysis (specific examples of inflationary generation mechanism for which the metric perturbations have also been calculated are found in~\cite{Anber:2006xt,maginf,barnaby,Barnaby:2011vw}). 

The simplest existing model for inflationary magnetogenesis consists in breaking conformal invariance through a coupling between the electromagnetic field and the inflaton $\varphi$ with an action of the form~\cite{turner,ratra}
\be
\label{actionem}
S=-\frac{1}{16\pi}\int d^4x \sqrt{-g}\, f^2(\varphi)F^{\mu \nu}F_{\mu \nu} + \, S_{\varphi,g} + \cdots \; ,
\ee
where $F_{\mu\nu}=A_{\nu,\mu}-A_{\mu,\nu}$ is the Faraday tensor, and $A_\nu$ is the electromagnetic 4-vector potential.  
The time evolution of the electromagnetic field depends on the coupling function $f(\varphi)$. We parametrize it directly as a function of conformal time obtained by inverting the background inflaton evolution $\bar\varphi(\eta)$. We consider the simple case 
\be
\label{f}
f(\eta)=f_1\left(\frac{\eta}{\eta_1}\right)^\gamma\; ,
\ee 
where we restrict $\gamma$ to the values $-2\leq\gamma\leq 2$. This ensures that the electromagnetic field 
remains subdominant and does not back react on the background expansion during inflation~\cite{yokoyama,subramanian}.
As shown in~\cite{maginf}, for super-horizon scales $|k\eta|<1$, the magnetic field power spectrum then scales as in Eq.~\eqref{PB}, and the magnetic field spectral index $n_B$ is related to $\gamma$ through $n_B=2\gamma +1$ for $\gamma<1/2$, and $n_B=3-2\gamma$ for $\gamma>1/2$. When $\gamma=-2$ the magnetic field is scale invariant ($n_B=-3$), and when $\gamma=1/2$, $n_B= 2$. 

In~\cite{maginf} we have computed the energy density and anisotropic stress of the electromagnetic field during inflation (here and in the following a sub- or super-script~$-$ denotes the quantities during inflation while $+$ indicates the radiation era directly after inflation). On super-horizon scales, $x=|k\eta|<1$, we have found
\bea
\ompm&\equiv&\sqrt{\frac{k^3 P_\Pi}{\bar{\rho}_\varphi^2}}=\frac{H^2}{3 m_P^2}C_\Pi(\gamma)\, x^\alpha\;, \label{omp} \\
\omem&\equiv&\sqrt{\frac{k^3 P_{\rm em}}{\bar{\rho}_\varphi^2}}=\frac{H^2}{3 m_P^2}C_{\rm em}(\gamma)\, x^\alpha\;, \label{omem}
\eea
where $P_\Pi(k, \eta)$ and $P_{\rm em}(k, \eta)$ are respectively the power spectra of the electromagnetic anisotropic stress and energy density as defined in~\cite{maginf}, $\bar{\rho}_\varphi$ is the background energy density during inflation, $H$ is the physical Hubble scale during inflation, and
\be
\alpha=\left\{\begin{array}{ll} 4+2\gamma=n_B+3 & \mbox{if} \quad -2\leq \gamma \leq -5/4\,, \\
3/2 & \mbox{if} \quad -5/4\leq \gamma \leq 5/4\,, \\
4-2\gamma=n_E+3 & \mbox{if} \quad 5/4\leq \gamma \leq 2 \,, \end{array} \right. 
\label{alphanB}
\ee
where $n_E$ denotes the spectral index of the electric field. The coefficients $C_\Pi(\gamma)$ and $C_{\rm em}(\gamma)$ are constants which depend on the value of $\gamma$. For the simple coupling of Eq.~\eqref{f} they are expected to be of order 1, but one could imagine specific models that would enhance or reduce their value. 
Note that an electric field is also generated by the coupling in Eq.~\eqref{f}. For $\gamma<-5/4$ the contribution of the electric field to the energy density $\Om^{-}_{\rm em}$ and to the anisotropic stress $\Om^{-}_\Pi$ is subdominant with respect to the magnetic one and can therefore be neglected. However, for $-5/4<\gamma<5/4$ the electric contribution is of the same order of the magnetic one and it enters in the coefficients $C_\Pi(\gamma)$ and $C_{\rm em}(\gamma)$. Finally, for $\gamma>5/4$ the electric field dominates and virtually no magnetic field is generated by the coupling under consideration. In this case, the electromagnetic anisotropic stress and energy density depend on the spectral index of the electric field 
$n_E$.

The energy density and anisotropic stress of the electromagnetic field act as sources for the curvature perturbation during inflation. In~\cite{maginf} we have calculated the comoving curvature perturbation and found that (c.f. Appendix~\ref{app:curvature} for a discussion)
\bea
\label{zetainf}
\zeta_-(x) &\simeq& \frac{H^2}{9\, m_P^2 \, \epsilon}  \Big[(\alpha-6) C_\eb(\ga)+
     \alpha \,C_\Pi(\ga) \Big]  \nonumber \\
&\times& \left\{\begin{array}{cc} -\log\big(x) &\mbox{if}~\al=0 ~ ~~ (|\ga|=2)\\
\al^{-1} & \mbox{if} ~\al\neq 0 \hspace{1.65cm}\end{array} \right.
\eea
where $\epsilon=(\HH^2-\dot\HH)/\HH^2$ is the slow roll parameter, $\HH = \dot a/a=aH$. The above expression for $\zeta_-(x)$ is valid at super-horizon scales $x<1$. Note that the pre factor $H^2/(\ep m_P^2)$ is of the order of the adiabatic power spectrum. Parametrically therefore, the power spectrum of this contribution is of the order of the square of the adiabatic one. However, the log can be large on large scales and also the term in square brackets in (\ref{zetainf}) may be substantially larger than 1.

We now study the impact of this inflationary magnetic mode on the CMB, and compare it with the passive and compensated modes.

\section{The curvature and metric perturbations after inflation}
\label{sec:psirec}

At the end of inflation the standard electromagnetic action is recovered. The magnetic field stops growing and is simply transferred to the radiation era, where it scales as a radiation component, i.e., $B^2\sim 1/a^4$. The electric field on the other hand is almost immediately dissipated due to the very high conductivity of the cosmic plasma~\cite{AE}. 

The energy density and anisotropic stress of the magnetic field continue to source the  curvature perturbation $\zeta$ and the metric potentials $\Phi$ and $\Psi$ also after inflation. The metric in the longitudinal gauge reads
\be
ds^2 = a^2\left[-(1+2\Phi)d\eta^2 + (1-2\Psi)d\bx^2\right] \,,
\ee
and $\zeta$ is related to $\Psi$ and $\Phi$ by
\be
\label{zetadef}
\zeta=\Psi+\frac{2}{3\HH(1+w)}\Big(\HH\Phi+\dot{\Psi} \Big)\; .
\ee
Combining Einstein's equations and the conservation equations one can write a second-order evolution equation for the curvature $\zeta$ valid after inflation, for the pre-recombination phase (we have set the total entropy perturbation in the fluids to zero and consider standard adiabatic initial conditions): 
\begin{align}
\label{evolzeta}
&\ddot{\zeta} +\left[2\HH+3\HH\big(w-c_s^2 \big)-2\frac{\dot{c}_s}{c_s} \right]\dot{\zeta}+c_s^2 k^2 \zeta=\\
& \frac{2w\HH}{1+w}\Bigg\{\left[\frac{2\dot{c}_s}{c_s}-\frac{\HH}{2}\big(1+3w+6 c_s^2 \big) \right] \Big(\ompp+\frac{R_\nu}{3}\pi_\nu\Big)\nonumber\\
&+\left[\frac{\HH}{4}\big(1-3c_s^2 \big)\big(1+3w\big)-\frac{\dot{c}_s}{c_s} \right]\ombp-\frac{R_\nu}{3}\dot{\pi}_\nu\Bigg\}\; ,\nonumber
\end{align}
where $\pi_\nu$ denotes the anisotropic stress of the neutrinos which starts to develop after neutrino decoupling, $R_\nu=\bar{\rho}_\nu/\bar{\rho}_{\rm rad}$, $c_s$ and $w$ are the background fluid sound speed and equation of state parameter and a dot denotes the derivative with respect to conformal time. 

$\ombp(k)$ is a function denoting the fractional energy density of the magnetic field at wavenumber $k$ and $\ompp(k)$ correspondingly denotes the fractional anisotropic stress. They are both constant in time, since they are normalized to the radiation energy density $\bar\rho_{\rm rad}$. We define the magnetic field energy density and anisotropic stress power spectra as
\begin{align}
\vev{\rho_B(\bk,\eta)\rho_B^*(\bq,\eta)}&=(2\pi)^3 P_{\rho_B}(k, \eta)\de(\bk-\bq)\,,\\
\vev{\Pi_B(\bk,\eta)\Pi_B^*(\bq,\eta)}&=(2\pi)^3 P_{\Pi_B}(k, \eta)\de(\bk-\bq)\,.
\end{align}
The quantities above have to be understood in terms of these spectra, as in Eqs.~\eqref{omp} and \eqref{omem}:
\be
 \ombp(k)\equiv \sqrt{\frac{k^3P_{\rho_B}}{\bar\rho_{\rm rad}^2}}~~~~~{\rm and}~~~~~\ompp(k)\equiv \sqrt{\frac{k^3P_{\Pi_B}}{\bar\rho_{\rm rad}^2}}\,.
 \label{PrhoB}
 \ee
If the magnetic field has its origin during inflation, the relation between $\omem$, $\ompm$ and $\ombp$, $\ompp$ is not completely trivial, since these quantities are not necessarily continuous at the transition from inflation to the radiation dominated era. During the radiation era many charged particles are present and the conductivity of the universe is high, so that the electric field is rapidly damped. As previously mentioned, in Ref.~\cite{maginf} we have found that the magnetic field dominates if $\ga<-5/4$ while the electric field dominates if $\ga>5/4$. If the electric field dominates  $\omem$, $\ompm$ during inflation, then $\ombp$, $\ompp$ can be significantly smaller than $\omem$, $\ompm$ because the electric field is dissipated after reheating. In the following, we shall however disregard this possibility, since we know that it cannot lead to significant magnetic fields. We therefore consider only $\ga\le 5/4$, corresponding to $n_B\le 2$ (note that the maximal $n_B$ is obtained for $\gamma=1/2$ where $n_B=3-2\gamma=2$). 

Eqs.~\eqref{omp} and~\eqref{omem} are valid at super-horizon scales $x<1$, for which the transition between the inflationary and the radiation dominated phase can be considered instantaneous. We assume that the transition happens at a time instant $\eta_*$, for which $\HH_{\rm inf}(\eta=-\eta_*) = \eta_*^{-1} = \HH_{\rm rad}(\eta=\eta_*)\equiv \HH_*=a_*H_*$, while the background equation of state parameter jumps from $w\simeq -1$ to $w=1/3$. Afterwards, we know that the components of the magnetic field energy momentum tensor decay in time as radiation. We therefore simply set for the fractional energy density and anisotropic stress
\bea
\ombp&\simeq&\frac{H_*^2}{3 m_P^2}C_{\rm em}(\gamma)\, x_*^\alpha\;, \label{ombprad}\\
\ompp&\simeq&\frac{H_*^2}{3 m_P^2} C_\Pi(\gamma)\, x_*^\alpha\; , \label{ompprad}
\eea
where $x_*=|k\eta_*|$ denotes the end of inflation, and we neglect the contribution from the electric field. 

Knowing $\ombp$ and $\ompp$, the evolution equation of the comoving curvature can be solved. Moreover, in order to evaluate the CMB anisotropies we want to determine also the metric perturbation $\Psi$, which is related to the curvature $\zeta$ by 
\bea
\label{evolpsi}
\lefteqn{\dot{\Psi}+\frac{\HH}{2}(5+3w)\Psi=\frac{3\HH}{2}(1+w)\zeta} \nonumber\\ 
& & \hspace*{2cm} +\, 9w\HH\left(\frac{\HH}{k}\right)^2 \Big(\ompp+\frac{R_\nu}{3}\pi_\nu\Big)\; .
\eea
This equation has been derived from Eq.~\eqref{zetadef} using the Einstein equation with indices $i\neq j$, 
\be
\Psi-\Phi=9w\left(\frac{\HH}{k}\right)^2\left[\ompp+\frac{R_\nu\,\pi_\nu}{3}\right]\,. 
\ee
To solve Eqs.~\eqref{evolzeta} and \eqref{evolpsi}, we further need to specify the background evolution and the neutrino anisotropic stress $\pi_\nu(k,\eta)$. Note that in this work we make the simplifying assumption that neutrinos are massless, since this does not affect our results substantially. For a  treatment of CMB magnetic field anisotropies (passive and compensated modes) with massive neutrinos see Ref.~\cite{shaw}.

\subsection{Before neutrino decoupling}

For $T\geq1$ MeV, neutrinos are coupled to the photons and have negligible anisotropic stress $\pi_\nu=0$. Moreover at these temperatures the universe is dominated by radiation and $w=c_s^2=1/3,\; \HH=1/\eta$. The evolution equations for $\zeta$ and $\Psi$ take the simple form
\be
\label{evolzetarad}
\zeta''+\frac{2}{x}\zeta'+\frac{\zeta}{3}=-\frac{\ompp}{x^2}\; ,
\ee
where a prime denotes a derivative with respect to $x=k\eta$; and
\be
\label{evolpsirad}
\Psi'+\frac{3}{x}\Psi=\frac{2}{x}\zeta+\frac{3}{x^3}\ompp\; .
\ee
Hence  the curvature $\zeta$ is sourced by the magnetic field anisotropic stress $\ompp$ at order $x^0$. The metric potential $\Psi$ is sourced by $\ompp$ at order $x^{-2}$ and by the curvature at order $x^0$.

Eq.~\eqref{evolzetarad} can easily be solved. The initial conditions for $\zeta$ are fixed by the matching conditions at the end of inflation. They insure the continuity of the induced 3-metric and the extrinsic curvature at the transition from the inflationary era to the radiation era \cite{muk_deruelle,maginf}. Without a magnetic field, this matching is trivial as all the relevant quantities $\zeta$, $\Psi$ and $\Phi =\Psi$ turn out to be continuous at the transition. In the presence of a magnetic field, however, these conditions imply that both $\zeta$ and $\Psi$ are continuous at the transition for large scales $x<1$, while $\Phi$ is not~\cite{maginf}. Also the time derivative of $\zeta$ is not continuous. To determine $\zeta'$ at the beginning of the radiation era, we need a relation between $\zeta$ and $\Psi$. Combining Einstein's equations with the derivative of $\zeta$ in Eq.~\eqref{zetadef} we find (c.f. Eq.~(16) of \cite{maginf})
\bea
\zeta'_{-}(x)&=&\frac{1}{\epsilon x}\left(x^2 \Psi_{-} +\ompm+\omem-9\Om^{-}_Q \right)\label{zetapmoins}\; ,\\
\zeta'_{+}(x)&=&-\frac{1}{2x}\left(\frac{x^2 \Psi_{+}}{3}+\ompp \right)\; .\label{zetapplus}
\eea 
In Eq.~\eqref{zetapmoins}, we keep only the lowest order terms in $\epsilon$. $\Om^{-}_Q$ represents the Poynting vector contribution; it is defined as in Eqs.~\eqref{omp} and \eqref{omem}, $\Om^{-}_Q=\sqrt{k^3P_Q/\bar\rho_\varphi^2}$, where $P_Q$ is the power spectrum of the quantity $i\HH\hat{k}^j{T^0}_j/k$, and ${T^0}_j$ is the Poynting vector component of the electromagnetic energy-momentum tensor \cite{maginf}. It is negligible when the electric field is subdominant ($\gamma<-5/4$) but of the same order of magnitude as $\ompm$ and $\omem$ for larger values of $\ga$.

Using that $\Psi$ is continuous at the transition, Eqs.~\eqref{zetapmoins} and \eqref{zetapplus} imply 
\be
\zeta'_{+}(x_*)=-\frac{\epsilon}{6}\zeta'_{-}(x_*) -\frac{\ompp}{2x_*}+\frac{\ompm+\omem-9\Om^{-}_Q}{6 x_*}\;,
\ee
where $x_*=|k\eta_*|$ denotes the end of inflation, and $\zeta'_{-}(x_*)$ is obtained by deriving Eq.~\eqref{zetainf} with respect to $x$. 
This condition, together with the continuity of $\zeta$ for which we know the solution (Eq.~\eqref{zetainf}), provides the initial conditions for Eq.~\eqref{evolzetarad} and allows us to find the curvature during the radiation era. We obtain
\bea
\hspace*{-3cm}\zeta_{+}&=& \zeta_{\rm inf}+\zeta_* +\ompp \log\left({\frac{\eta_*}{\eta}}\right)  
\nonumber\\   \label{e:zeta+}
& & \qquad +\left(1-\frac{\eta_*}{\eta} \right)\left[\frac{\ompp}{2}+\frac{\Om_*}{6} \right]\,,
\eea
where $\zeta_*\equiv \zeta_{-}(x_*)$ [c.f. Eq.~\eqref{zetainf}] and
\be
\label{omstar}
\Om_*=\frac{1}{3}\left\{\begin{array}{ll} \ompm+\omem-9\Om^{-}_Q & \mbox{if} \;\alpha \neq 0\\
 \ompm+\omem-9\Om^{-}_Q-\frac{\epsilon\, \zeta_*}{\log\left(x_*\right)} & \mbox{if}\; \alpha= 0\,.\end{array} \right.
\ee

There are four distinct contributions to $\zeta_{+}$:
\begin{enumerate}[(i)]
\item $\zeta_{\rm inf}$ is the standard scalar adiabatic mode. 
\item The contribution proportional to $\ompp$ is a dynamical mode generated by the magnetic field anisotropic stress \textit{after} the end of inflation. This mode has already been computed in~\cite{magcaus} and in~\cite{shaw} where it is called the passive magnetic mode. It vanishes at the transition, $\eta=\eta_*$.
\item $\zeta_*\equiv \zeta_{-}(x_*)$ is a new contribution, computed for the first time in~\cite{maginf}. It is directly transmitted from inflation to insure the continuity of the curvature.  It is therefore a remnant of the inflationary period and is absent if the magnetic field is generated causally after inflation~\cite{magcaus}. This {\em inflationary magnetic mode} is constant in time. It is not affected by the behavior of the magnetic field after inflation.
\item The term proportional to $\Om_*$ is also a new contribution, generated by the continuity of $\Psi$ at the transition. It is however negligible with respect to $\zeta_*$. First of all, it is one order higher in the slow roll parameter $\epsilon\ll 1$ (therefore, it is of the same order in $\ep$ as other terms which we have neglected in the derivation of expression~(\ref{e:zeta+})). Moreover, it is further reduced by other factors. From Eqs.~\eqref{zetainf} and \eqref{omstar} we see that the first part of $\Om_*$ is of the order $\ompm(x_*)\sim \epsilon\, x_*^\alpha \zeta_*\ll\zeta_*$ since $x_*\ll1$, $\alpha\geq 0$ and $\epsilon\sim0.01$ at the end of inflation. For $\alpha=0$ it contains a part $(\epsilon\,/\log x_*)\,\zeta_*$, with $|\log x_*|\gg 1$. Note that the amplitude of the first part of $\Om_*$ is in principle of the same order as $\ompp$. However, it is a non-dynamical component arising directly from inflation, of the same nature of the inflationary magnetic mode $\zeta_*$: we consistently compare it only to this latter.
\end{enumerate}
We neglect the subdominant $\Om_*$--contribution and the decaying mode so that well into the radiation era, for $\eta\gg \eta_*$, the curvature takes the simple form ($\eta_\nu$ denotes the time of neutrino decoupling)
\be
\label{zetasolrad}
\zeta\simeq \zeta_{\rm inf}+\zeta_*+ \left[ \frac{1}{2}+\log\left({\frac{\eta_*}{\eta}}\right) \right]\omp\,, ~~~~~\eta_*\ll \eta\leq \eta_\nu 
\ee
where here and in the following we drop the superscript~$+$ for simplicity. Here, $\omp$ is the magnetic anisotropic stress ratio in the radiation era. 

Inserting the solution~\eqref{zetasolrad} into Eq.~\eqref{evolpsirad} and integrating, we find the Bardeen potential during the radiation era
\bea
\label{psisolrad}
\Psi \simeq \frac{2}{3}\zetainf +\frac{2}{3}\zeta_* +\frac{3\omp}{x^2}+\left[\frac{5}{9}+\frac{2}{3} \log\left({\frac{\eta_*}{\eta}}\right) \right]\omp\, , \\
& & \hspace*{-3cm} \eta_*\ll\eta\leq \eta_\nu\,. \nonumber
\eea
The inflationary magnetic mode, $\zeta_*$, generated by the electromagnetic field during inflation, contributes to the Bardeen potential in the same way as the standard inflationary mode, $\zeta_{\rm inf}$. As the large scale CMB anisotropies are determined by the Bardeen potentials\footnote{Note that the CMB temperature fluctuation $\Delta T/T$ from the Sachs Wolfe effect can be determined directly by the curvature perturbation $\zeta$ only in the standard adiabatic case for which $\Phi=\Psi$ and the potentials are almost constant during the matter era. In the presence of non-zero anisotropic stresses at large scales, i.e. when a magnetic field is present, $\Delta T/T$ depends on both Bardeen potentials, as given in Eq.~\eqref{SWV}.}, we expect the inflationary magnetic mode to generate CMB temperature anisotropies in a similar way as the standard inflationary term, although with a smaller amplitude (c.f. Eq.~\eqref{zetainf}). To determine the CMB temperature anisotropies, we need to evolve the metric perturbations further, until recombination, accounting for neutrino decoupling.

\subsection{After neutrino decoupling}

To evolve the curvature and metric potentials until recombination and evaluate the effect of the inflationary magnetic mode $\zeta_*$ on the temperature power spectrum we need to take into account that at a temperature $T_\nu\sim 1$~MeV neutrinos decouple from the electron-positron-photon plasma and start to free stream. They develop an anisotropic stress that acts as a source for the curvature and metric perturbations in~Eqs.~\eqref{evolzeta} and~\eqref{evolpsi}. Moreover, at~$T_{\rm eq}\simeq 0.73$~eV the universe evolves from radiation domination to matter domination. The equation of state $w$ and the sound speed $c_s^2$ depart from their radiation value. As a result, not only the anisotropic stress but also the magnetic energy density, $\omb$, source the curvature perturbation, see Eq.~\eqref{evolzeta}.

\begin{figure}
\includegraphics{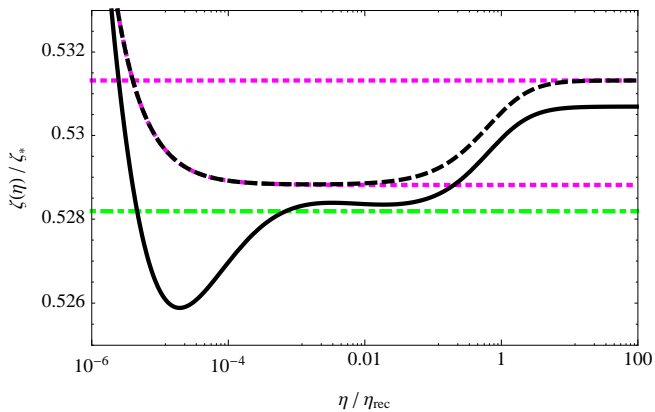}
\caption{The comoving curvature $\zeta$ generated by an inflationary magnetic field. Solid, black: the correct numerical solution with the true $\pi_\nu$. Dashed, black: the approximated solution calculated from approximation \eqref{pinuapp}. Dotted, magenta: the asymptotic expressions given in Eqs.~\eqref{zetasmall} (lower line) and \eqref{zetabig} (upper line). Dash-dotted, green: the approximation given in Eq.~(86) of Ref.~\cite{shaw}. For this figure, we have set $\omb=\ompp\simeq \ompm(x_*)\simeq \epsilon \,\zeta_*\simeq 0.01\,\zeta_*$, $\eta_*/\eta_\nu=10^{-22}$, $\eta_\nu/\eta_{\rm rec}=10^{-6}$ (the plot shows $\zeta$ after neutrino decoupling). Note also the vertical axes: the differences are small, on the level of 1\%.}
\label{fig:curv}
\end{figure}

As studied in detail in~\cite{magcaus} (see also~\cite{shaw, Kojima:2009gw,const}), once neutrinos decouple, their anisotropic stress quickly adjusts at large scales to compensate the one of the magnetic field. The precise dynamical behaviour of $\pi_\nu$ can be determined by combining the Boltzmann hierarchy with the Einstein's equations. It is well approximated by~\cite{magcaus}
\be
\pi_\nu(\eta>\eta_\nu)\simeq -\frac{3\omp}{R_\nu}\left(1-\frac{\eta_\nu^2}{\eta^2} \right) + \mathcal{O}(k\eta)^2\,.
\label{pinuapp}
\ee
This compensation affects the subsequent evolution of $\zeta$ and $\Psi$. Indeed, after compensation the magnetic anisotropic stress $\omp$ does not source the curvature perturbation anymore. This stops the logarithmic growth of $\zeta$ at $\eta\simeq \eta_\nu$ (c.f Eq.~\eqref{zetasolrad}). The only remaining source term in Eq.~\eqref{evolzeta} is then the magnetic energy density $\omb$ with a pre-factor which vanishes in the radiation era. The inflationary magnetic mode $\zeta_*$ is not affected by this compensation and it simply goes through neutrino decoupling without alteration. The full solution for $\zeta$ is rather complicated, hence we only write its behaviour in the limits $\eta_\nu\ll\eta\ll \eta_{\rm eq}$ and $\eta\gg \eta_{\rm eq}$, details can be found in Ref.~\cite{magcaus}:
\bea
\zeta&\simeq &\zetainf+\zeta_*+\left[\frac{\eta_\nu}{\eta}-\frac{1}{2}+\log\left(\frac{\eta_*}{\eta_\nu}\right) \right]\omp\,, \label{zetasmall}\\
& & \hspace*{2.5cm} \eta_\nu\ll\eta\ll \eta_{\rm eq}\,,\nonumber\\
& & \nonumber \\
\zeta&\simeq &\zetainf+\zeta_*+\frac{\omb}{4}+\left[-\frac{1}{2}+\log\left(\frac{\eta_*}{\eta_\nu}\right) \right]\omp\,, \label{zetabig}\\
& & \hspace*{3.5cm} \eta\gg \eta_{\rm eq} \,.\nonumber
\eea
The analytical solution for $\zeta$, together with the exact solution obtained by integrating Eq.~\eqref{evolzeta} numerically  with the correct expression for the neutrino anisotropic stress instead of approximation \eqref{pinuapp} (c.f. Appendix B of~\cite{magcaus}), are shown in Fig.~\ref{fig:curv}.  Approximation \eqref{pinuapp} adjusts to the value $-3\omp / R_\nu$ in a way which differs somewhat from how the true $\pi_\nu(k,\eta)$ adjusts itself to the same value. This is reflected in the curvature, leaving an offset between the approximated solution and the true one even at late times. Since this offset is much smaller than the amplitude of the passive mode in $\zeta$, we neglect it and keep using the analytical solution in the following. 

The metric perturbation $\Psi$ is also affected by the compensation, that cancels the anisotropic source term scaling as $(\HH/k)^2$ in Eq.~\eqref{evolpsi}. As a result, the term $3\omp/x^2$ in Eq.~\eqref{psisolrad} cancels. Hence the metric perturbations $\Psi$ and $\Phi$ contribute to the large scale CMB anisotropies only at next-to-leading order $(k\eta)^0$. This has already been found in \cite{magcaus} for the case of a magnetic field generated by a causal process (e.g. a primordial phase transition). 

Solving Eq.~\eqref{evolpsi} at order $(k\eta)^0$ requires us to know the evolution of the neutrino anisotropic stress at order $(k\eta)^2$. Combining the Boltzmann hierarchy with Einstein's and the conservation equations in the radiation era gives the following expression for $\pi_\nu$ at next-to-leading order 
\bea
\label{pinunext}
&\pi_\nu(k,\eta)&=\Bigg\{\frac{4}{15+4R_\nu}\left[\zeta_*+\log\left(\frac{\eta_*}{\eta_\nu}\right)\omp\right]+\\
& &\frac{165\omp-42R_\nu\omb}{14R_\nu (15 + 4 R_\nu)}\Bigg\}k^2\eta(\eta-\eta_\nu) ~~~{\rm at}~\mathcal{O}(k\eta)^2\, .\nonumber
\eea
This solution is strictly valid only during the radiation era and it obtains corrections at the transition into the matter era. In our analytical approximation we use it, however, until recombination since analytically it is not possible to derive the exact solution. We do not expect those corrections to change the result significantly.  In the numerical calculations we just use it for the initial conditions which are in the radiation era but after neutrino decoupling.

Inserting the solutions for $\pi_\nu$ and $\zeta$ into Eq.~\eqref{evolpsi} we find $\Psi$ after neutrino decoupling (the initial condition is given by Eq.~\eqref{psisolrad}). Here we only write the inflationary magnetic contribution $\zeta_*$ to $\Psi$  :
\bea
\Psi^{\rm im}&=& \frac{2(5+2R_\nu)}{15+4 R_\nu}\zeta_* \;,  \quad \eta \ll \eta_{\rm eq} \label{psinu}\\
\Psi^{\rm im}&=&\frac{3}{5}\zeta_*\;, \hspace{1.9cm} \eta\gg \eta_{\rm eq}\,.
\eea 
The inflationary magnetic mode is affected by neutrino decoupling. It evolves from $\Psi^{\rm im}=2\zeta_*/3$ before decoupling (see Eq.~\eqref{psisolrad}) to expression~\eqref{psinu} after decoupling. This behaviour is like the one of the standard adiabatic curvature perturbation $\zetainf$ that is also affected by the anisotropic stress of the neutrino. Then, well into the matter era, $\Psi^{\rm im}$ tends to $3\zeta_*/5$. In addition to the inflationary magnetic mode, $\Psi$ contains also the passive and compensated modes already calculated in~\cite{magcaus, shaw, finelli}.

\section{The Sachs Wolfe effect, analytic approximation }
\label{sec:SW}

We are now able to calculate the magnetic field contribution to the CMB anisotropy. In this section we present an analytical estimate of the Sachs Wolfe amplitude in order to gain insight into the relative amplitudes of the terms and their scaling. Following~\cite{magcaus} we express the Sachs Wolfe as
\bea
\label{SWV}
\frac{\De T}{T}&=&\frac{D_{g\,\gamma}}{4} +\Psi+\Phi\\
 &=&\frac{1}{k}\dot{V}_\gamma-\frac{\omb-2\omp}{4(1-R_\nu)}\; ,\nonumber
\eea
where the last term after the second equality sign is the contribution from the Lorentz force. 
The photon velocity $V_\gamma$ obeys the second order differential equation
\be
\label{Vgamma}
\ddot{V}_\gamma+\frac{k^2}{3}V_\gamma=k(\dot{\Phi}+\dot{\Psi})\; .
\ee
Since we are interested in the large-scale solution, we can neglect the $k^2$-term in the above equation, so that
\be
\label{Vsol}
\dot{V}_\gamma=k(\Phi+\Psi)+c\; .
\ee
 In Eq.~\eqref{SWV} we have used $\dot{V}_\gamma =k[\Psi+\Phi + D_{g\,\gamma}/4 +(\omb-2\omp)/4/(1-R_\nu)]$.
This implies that on large scales  $D_{g\,\ga} +(\omb-2\omp)/4/(1-R_\nu) =c/k$ is constant. However, analytically there is no way to determine directly the quantity $D_{g\,\ga}$ other than via the evolution equation of $V_\ga$. The constant of integration $c$ can in fact be determined from the initial conditions, using that at neutrino decoupling $\dot{V}_\gamma(\eta_\nu)=\dot{V}_\nu(\eta_\nu)=\dot{V}(\eta_\nu)$. Up to neutrino decoupling, photons and neutrinos behave indeed as a single fluid and share the same velocity. Inserting the solution~\eqref{Vsol} into Eq.~\eqref{SWV} we find at $\eta=\eta_{\rm rec}$
\bea
\frac{\De T}{T}&\simeq& \frac{505 + 108 R_\nu}{135(15 + 4 R_\nu)}\left[\zetainf+\zeta_*+ \omp\log\left(\frac{\eta_*}{\eta_\nu} \right) \right]\nonumber \\
&+&\frac{(765 + 244 R_\nu)}{270 (15 + 4 R_\nu)}\omb -\frac{184005 + 48188 R_\nu}{ 5670 (15 + 4 R_\nu)}\omp\nonumber\\
&-&\frac{\omb-2\omp}{4(1-R_\nu)}\; .
\label{SWstar}
\eea
The first term is the standard adiabatic contribution to the Sachs Wolfe term. The second term is our inflationary magnetic mode, equivalent to the adiabatic one. The third term is the contribution from the passive mode. Note that this part obtains corrections if one relaxes the assumption of Eq.~\eqref{pinuapp}. The way in which the neutrino anisotropic stress adjusts to the compensation value~$-3\Omega_\Pi/R_\nu$  in fact introduces a correction of the order $\Omega_\Pi$ in the passive mode (as shown in Fig.~\ref{fig:curv} for the curvature perturbation). This correction is however negligible with respect to the logarithmic term. The second line represents the contribution from the compensated mode~\footnote{The Sachs Wolfe term from the compensated mode is slightly different from the one given in Eq.~(6.11) of \cite{maginf}, due to the fact that \cite{maginf} uses a different initial expression for $\pi_\nu$ at next-to-leading order.}. Finally, the third line contains the contribution from the Lorentz force.

As the Bardeen potentials from the compensated mode are not constant in time, there is also an integrated Sachs Wolfe term which is not accounted for in this analytical approximation. 

The inflationary magnetic mode $\zeta_*$ contributes to the Sachs Wolfe effect differently from the passive and compensated modes:
 the temperature angular power spectrum of the passive and compensated modes depends on the magnetic field spectral index, that determines the k-dependence of $\omb$ and $\omp$ as $(k\eta_*)^\al$ (c.f. Eqs.~\eqref{ombprad} and~\eqref{ompprad}). 
On the other hand, since $\zeta_*$ is independent of $k$, i.e. scale invariant (up to a possible log correction if $\al=0$) for any spectral index $n_B$ (c.f. Eq.~\eqref{zetainf}), its impact on the temperature power spectrum is always scale invariant like the inflationary one: $\ell(\ell+1)C_\ell\propto\ell^0$. Hence, through this inflationary magnetic mode, even a blue magnetic field can leave a detectable imprint on the CMB at large scales, if it has a sufficient amplitude. 

In the following we compare the amplitude and scaling of the inflationary magnetic mode with the passive and compensated modes, using both analytical estimates and numerical evaluations: these latter are obtained using the modified CAMB code of Ref.~\cite{shaw}.

\section{The angular power spectrum}
\label{sec:angular}

Let us estimate the CMB anisotropy generated by the different modes. We use the Fourier convention
\be
f(\bk)=\int d^3x\, e^{-i \bx\cdot\bk}f(\bx)\; \label{fourier},
\ee
so that, as shown in \cite{magcaus} (for details see~\cite{mybook}), the temperature power spectrum from the Sachs Wolfe effect at large scales can be approximated by 
\bea
C_\ell \simeq \frac{2}{\pi}\int_0^\infty dk\,k^2 g^2(\eta_{\rm rec}) j_\ell^2(k,\eta_0)  P_\frac{\Delta T}{T}(k, \eta_{\rm rec})\, ,
\eea
where $g(\eta_{\rm rec})$ is the visibility function, $j_\ell(k,\eta_0)$ the spherical Bessel function and $P_\frac{\Delta T}{T}(k, \eta_{\rm rec})$ the spectrum of the temperature perturbation, the square of the terms in Eq.~\eqref{SWstar}. 
The amplitude of the inflationary magnetic mode is given by
\bea
\zeta^2_* &=&k^3 P_{\zeta_*}= \left(\frac{H_*^2}{m_P^2\epsilon}\right)^2\frac{1}{81}  \Big[(\alpha-6) C_\eb+
     \alpha \,C_\Pi \Big]^2  \nonumber \\
&\times& \left\{\begin{array}{cc} \log^2\big(x_*) &\mbox{if}~\al=0\\
1/\alpha^2 & \mbox{if} ~\al\neq 0\end{array} \right. \label{zetastarspectrum}
\eea
so that the CMB power spectrum at large scales is effectively flat, just as the inflationary one:
\bea
\frac{\ell(\ell+1)C_\ell^{\rm im}}{2\pi}  \simeq \frac{g^2(\eta_{\rm rec})}{2\pi^2} \left[ \frac{505 + 108 R_\nu}{135(15 + 4 R_\nu)} \right]^2\!\! 
\left(\frac{H_*^2}{m_P^2\epsilon}\right)^2\times  \nonumber \\ 
 \frac{1}{81}  \Big[(\alpha-6) C_\eb+ \alpha \,C_\Pi \Big]^2  
 \!\!\left\{\begin{array}{cc} \log^2\big(\eta_*/\eta_0\big) &\mbox{if}~\al=0\\
1/\alpha^{2} & \mbox{if} ~\al\neq 0\end{array} \right. \hspace*{0.4cm}  \label{e:Clim}
\eea

For the passive mode, the CMB power spectrum depends instead on the magnetic field spectral index. One has to consider two cases: for $-3<n_B<-3/2$, the anisotropic stress power spectrum behaves as $k^3P_\Pi\propto k^{2n_B+6}$ (c.f. Eqs.~\eqref{ompprad} and \eqref{alphanB}). To evaluate the integral of the Bessel function, one can use approximations (A3) and (A4) of \cite{Caprini:2009vk}. Restricting for example to the case $n_B<-2$, using Eq.~\eqref{ompprad} one finds
\begin{align}
\frac{\ell(\ell+1)C_\ell^{\rm pass}}{2\pi} \simeq \frac{g^2(\eta_{\rm rec})}{4 \pi^{3/2}} \left[ \frac{505 + 108 R_\nu}{135(15 + 4 R_\nu)} \right]^2 \log^2&\left(\frac{\eta_*}{\eta_\nu}\right) \nonumber\\
\times\left(\frac{H_*^2}{3m_P^2}\right)^2  {C_\Pi}^2(\gamma)  \left( \frac{\eta_*}{\eta_0}\right)^{2n_B+6}\frac{\Gamma[-n_B-2]}{\Gamma[-n_B-3/2]} \,\,& \ell^{2n_B+6} \nonumber \\
\hspace{-2cm} {\rm for}~~n_B <-2\,.  \label{Clpassivered}
\end{align}
If the magnetic spectral index is not scale invariant $n_B\not\simeq  -3$, the amplitude of the CMB spectrum is severely suppressed by the factor $(\eta_*/\eta_0)^{2n_B+6}$. 
For $n_B>-2$, the above integral diverges in the UV and we have to introduce the cutoff $k<k_D$ (c.f. Eq.~\eqref{PB}). If $n_B>-3/2$, the anisotropic stress power spectrum is flat $k^3P_\Pi\propto k^{3}$. One can use approximation (A2) of \cite{Caprini:2009vk}. The value of the integral of the Bessel function depends on the upper cutoff of the magnetic spectrum $k_D$, and in the limit $k_D\eta_0\gg 1$~\cite{Caprini:2009vk}, the CMB spectrum increases as $\ell^2$ for any value of the magnetic power spectrum: 
\begin{align}
\frac{\ell(\ell+1)C_\ell^{\rm pass}}{2\pi}& \simeq \frac{g^2(\eta_{\rm rec})}{2\pi^2} \left[ \frac{505 + 108 R_\nu}{135(15 + 4 R_\nu)} \right]^2 \log^2\left(\frac{\eta_*}{\eta_\nu}\right) \nonumber \\
&\times\left(\frac{H_*^2}{3m_P^2}\right)^2  {C_\Pi}^2(\gamma)  \left( \frac{\eta_*}{\eta_0}\right)^{3} (k_D\eta_0) \,\ell^2 \nonumber\\
&\hspace{3cm}{\rm for}~~n_B >-3/2\,. \label{Clpassiveblue}
\end{align}
Also in this case the result is severely suppressed by the factor $(\eta_*/\eta_0)^3$ (note that $k_D\eta_0\ll (\eta_0/\eta_*)^3$).

The contributions from the compensated mode and the Lorentz force in Eq.~\eqref{SWstar} are also proportional to $\Omega_\Pi$ and $\Omega_{\rm B}$. The Sachs Wolfe term from these modes has therefore the same $\ell$-dependence as the Sachs Wolfe term from the passive mode, shown in Eqs.~\eqref{Clpassivered} and~\eqref{Clpassiveblue}. The amplitude is reduced due to the absence of the logarithmic term in Eq.~\eqref{SWstar}. However, in addition to the Sachs Wolfe effect, we expect also a significant integrated Sachs Wolfe at large scale, since the compensated mode and the Lorentz force generate non-constant metric potentials $\Phi$ and $\Psi$. Our approximations, that only take into account the Sachs Wolfe term, are consequently not expected to be accurate for these contributions (therefore we do not show them here).

From these rough analytic estimates we infer that the contribution of the inflationary magnetic mode to the CMB anisotropies always dominates over the passive and the compensated modes. Comparing Eqs.~\eqref{ompprad} and \eqref{zetastarspectrum}, one  sees that the amplitude of a given $k$ of the passive mode is suppressed by a factor $\epsilon^2$ and $x_*^{2\alpha}\ll1$ with respect to the inflationary magnetic mode. In the $C_\ell$ spectrum the second suppression is converted to a factor $(\eta_*/\eta_0)^{2\al}$. The suppression is stronger for bluer magnetic fields with a larger $\alpha$. Assuming that the magnetic anisotropic stress and energy density are perfectly anti-correlated, $C_\eb(\gamma)=-C_\Pi(\gamma)$ (as suggested by the numerical analysis of~\cite{shaw, finelli}), and restricting to the case $\gamma<-5/4$, so that the magnetic field dominates and its energy density and anisotropic stress are continuous at the transition to the radiation era, one finds 
\bea
\frac{C_\ell^{\rm pass}}{C_\ell^{\rm im}} \simeq 
\epsilon^2 \left[\log^2\left( \frac{\eta_*}{\eta_\nu}\right)\right] \left( \frac{\eta_*}{\eta_0}\right)^{2n_B+6} \frac{\Gamma[-n_B-2]}{\Gamma[-n_B-3/2]} 
\nonumber\\
 \ell^{2n_B+6}  \left\{\begin{array}{cc} 1/\log^2\big( \eta_*/\eta_0 \big) & \mbox{if}~n_B\rightarrow -3 \\
(n_B+3)^{2} \, &  \mbox{if} ~-3<n_B<-2\end{array} \right. \nonumber
\eea
It appears that $C_\ell^{\rm pass}\ll C_\ell^{\rm im}$. The same applies if $n_B>-3/2$, c.f. Eqs.~\eqref{e:Clim} and \eqref{Clpassiveblue}. 

Note that this analysis applies to the case of a magnetic field entirely generated during inflation and transmitted to the radiation era without amplification. However, processes during reheating may lead to significant amplification of the magnetic field~\cite{ruthreview}. Moreover, phase transitions in the early universe are expected to generate primordial magnetic fields as well~\cite{xx}. Such processes would amplify the passive and compensated mode, but they are not expected to modify the inflationary magnetic mode. They may consequently change the above ratio between the passive mode and the inflationary magnetic mode.

\begin{figure}[!ht]
\centerline{\epsfig{figure=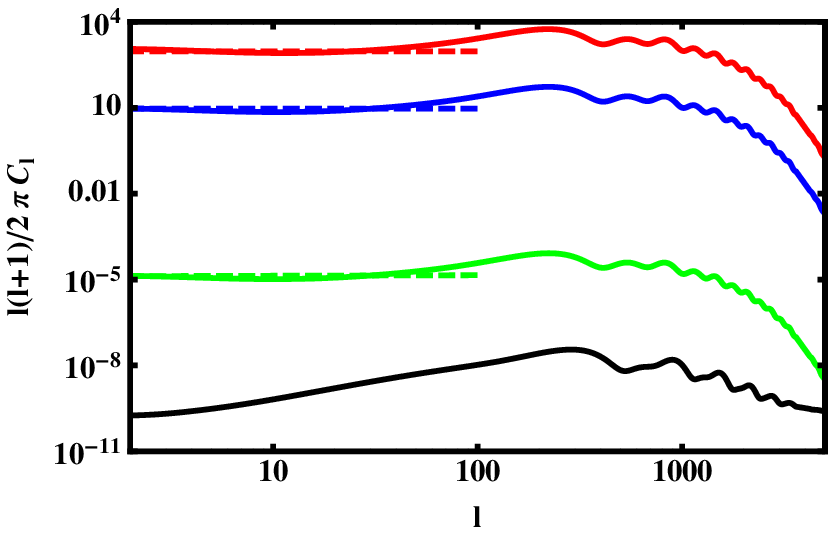,height=5.2cm}}
\centerline{\epsfig{figure=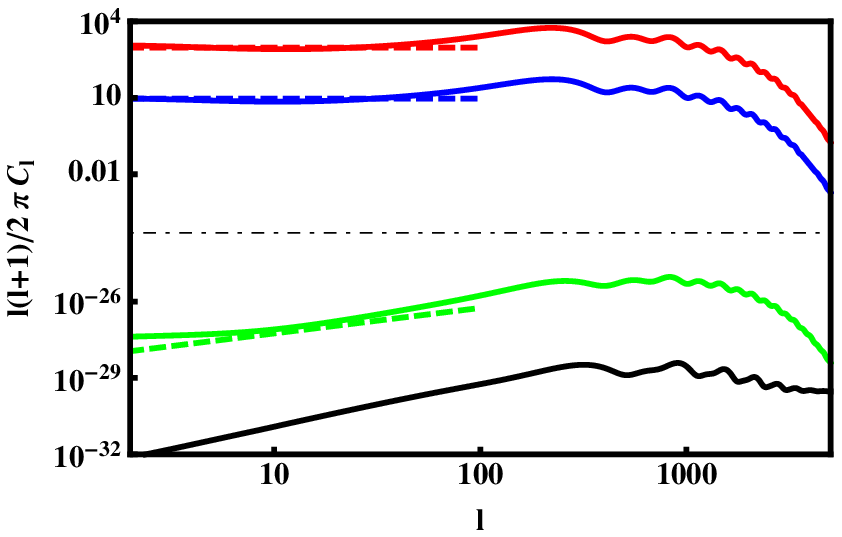,height=5.2cm}}
\centerline{\epsfig{figure=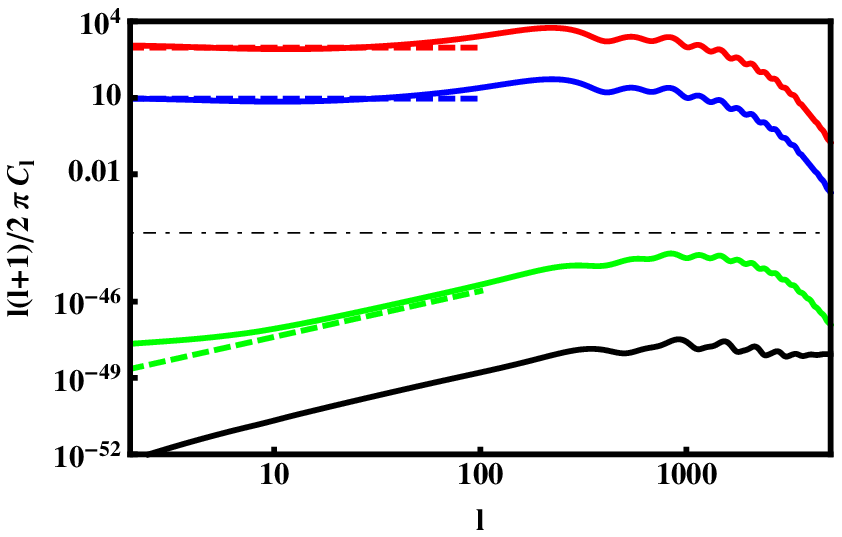,height=5.2cm}}
\caption{ \label{fig:cl} Temperature angular power spectrum (in $\mu {\rm K}^2$) for a magnetic field with spectral index : $n_B=-2.99$ (upper panel), $n_B=-5/2$ (middle panel) and $n_B=-2.1$ (bottom panel).  For each case we plot (from top to bottom) the standard adiabatic mode (red), the inflationary magnetic mode (blue), the passive mode (green) and the compensated mode with Lorentz force (black). The solid lines are the numerical results from CAMB and the dashed lines are our analytical approximations to the Sachs Wolfe effect given in Eqs.~\eqref{e:Clim} to \eqref{Clpassiveblue}, valid at low $\ell$. Note that in the middle and bottom panel we do not show the 24 (respectively 44) orders of magnitude between the inflationary magnetic mode and the passive mode.}
\end{figure}

In Fig.~\ref{fig:cl} we show some examples of CMB spectra for the standard inflationary mode and the magnetic modes, derived using the modified CAMB code of~\cite{shaw}, and we compare them with our analytical approximations at large scale. The amplitude of the standard adiabatic mode is
\be
\zeta^2_{\rm inf}=k^3P_{\zetainf}=\frac{2\pi H^2_*}{m_P^2\epsilon}\simeq 2\pi^2\cdot 2.1\times10^{-9}\;,\label{zetainfvalue}
\ee
where the factor $2\pi^2$ accounts for our Fourier convention~\eqref{fourier} and power spectrum convention~\eqref{mpowerspectrum} that are different from those used in CAMB.
From Eq.~\eqref{zetastarspectrum} we see that the inflationary magnetic mode is suppressed by a factor $H^2_*/(m_P^2\epsilon)\simeq 6\times10^{-9}$ with respect to the standard adiabatic mode of Eq.~\eqref{zetainfvalue}. In order to leave an impact on the CMB, the coefficients $C_\Pi^2/\alpha^2$ and  $C_\eb^2/\alpha^2$ need therefore to be large. In Fig.~\ref{fig:cl} we choose $C_\Pi=-C_\eb$ such that $\zeta_*^2\sim 0.01\cdot\zeta^2_{\rm inf}$. We consider three different cases:
\begin{itemize}
\item $n_B=-2.99$, so that $\alpha=0.01$, and $C_\Pi=46$\;,
\item $n_B=-5/2$, so that $\alpha=1/2$ and $C_\Pi=2315$\;,
\item $n_B=-2.1$, so that $\alpha=0.9$ and $C_\Pi=4166$\;.
\end{itemize}
Clearly, blue spectra require large, unphysical  values of $C_\Pi$ in order to leave a $\sim1\%$~impact on the CMB. This requires fine-tuning of the generation mechanism: our simple model produces indeed $C_\Pi\sim 20$ for $n_B=-2.99$ and values of order unity for $n_B=-5/2$ and $n_B=-2.1$. In Fig.~\ref{fig:cl} we see that the inflationary magnetic mode dominates over the passive and compensated modes at all multipoles by several orders of magnitude. It therefore leads to much stronger constraints on the primordial amplitude of the magnetic field than the passive and compensated modes considered in~\cite{shaw, finelli}. We also compare the numerical angular power spectrum with our analytical expressions Eqs.~\eqref{e:Clim} and \eqref{Clpassivered} (dashed line), which provide a good approximation at large scale $\ell \lesssim 100$. Note that for these approximations we have chosen $\epsilon\simeq 0.01$ and $\eta_*$ is determined from Eq.~\eqref{zetainfvalue} so that $\eta_*/\eta_0\simeq 3.1\times 10^{-28}$.

\section{Discussion}
\label{sec:discussion}

We have demonstrated that the inflationary magnetic mode always dominates the passive and compensated ones. It should therefore be taken into account when constraining primordial magnetism with the CMB. A full MCMC analysis is beyond the scope of this paper, but it is possible to predict analytically the constraints on the late time magnetic field that would result from the inflationary mode. In order to do so, we want to relate the amplitude of $\zeta_*$ to the magnetic field amplitude $B_\la$ and the spectral index $n_B$. 

Under the hypothesis that the energy density and the anisotropic stress of the magnetic field are perfectly anti-correlated $C_\eb(\gamma)=-C_\Pi(\gamma)$, $\zeta_*$ can be directly related to the magnetic field energy density $\omb$, c.f Eqs.~\eqref{zetastarspectrum} and~\eqref{ombprad}. The magnetic energy density power spectrum at the beginning of the radiation era is 
\be 
k^3 P_{\rho_B}(k,\eta_*)=[3 m_P^2 H_*^2\, \omb]^2\,,
\label{PrhoBeta*}
\ee 
obtained from Eq.~\eqref{PrhoB} where we have set the energy density at the beginning of the radiation era $\bar\rho_{\rm rad}(\eta_*)\equiv \rho_*=3H_*^2m_P^2=g_{\rm eff}^*a_{\rm SB} T_*^4$. The energy density power spectrum today becomes then $P_{\rho_B}(k,\eta_0)=(a_*/a_0)^8 P_{\rho_B}(k,\eta_*)$, which can be expressed in terms of $B_\lambda$. In order to do this, we use the definitions and the approximated spectra given in Sec. II of Ref.~\cite{Caprini:2009vk}, which read:
\bea
P_{\rho_B}(k,\eta_0) \simeq \left\{
\begin{array}{ll}
\frac{A^2_B(\eta_0)\,k_D^{2n_B+3}}{32\pi^4(2n_B+3)}\,, & {\rm if}~n_B<-\frac{3}{2} \\
\\
 \frac{3A^2_B(\eta_0)\,n_B\,k^{2n_B+3} }{128\pi^4(2n_B+3)(n_B+3)}\,, & {\rm if}~n_B>-\frac{3}{2} \\
\end{array}\right.
\eea
where $A_B(\eta_0)$ is related to $B_\lambda$ through Eq.~\eqref{Bla}. Together with Eq.~\eqref{PrhoBeta*}, we have then all the ingredients to relate $\omb$ and therefore $C_\Pi(\gamma)$ to $B_\lambda$ and $n_B$, and to substitute it into Eq.~\eqref{zetastarspectrum}. We finally obtain ($T_0$ denotes the temperature today):
\bea
&&\zeta_*= \frac{B_\lambda^2}{\rho_*}\, \frac{1}{\epsilon}\, \left(\frac{T_*}{T_0}\right)^4 \label{zetaBlambda}\\
&&\left\{\begin{array}{ll}
\sqrt{\frac{3n_B}{8\Gamma^2[\frac{n_B+3}{2}] (n_B+3)^3(2n_B+3)}} \left(\frac{\lambda}{\eta_*}\right)^{n_B+3}\,, & {\rm if}~ n_B<-\frac{3}{2} \\
\\
\frac{2}{\sqrt{32}\Gamma[\frac{n_B+3}{2}] \sqrt{2n_B+3}} \left(\frac{\lambda}{\eta_*}\right)^{3/2} (k_D\lambda)^{n_B+\frac{3}{2}}\,, & {\rm if} ~ n_B>-\frac{3}{2} 
\end{array}
\nonumber
\right.
\eea

\begin{figure}[ht]
\centerline{\epsfig{figure=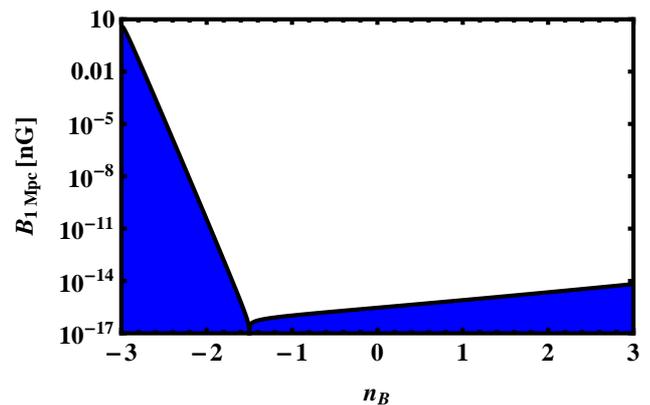,height=5.3cm}}
\caption{ \label{fig:Blambda} Upper bound on the magnetic field amplitude today smoothed on a scale of $1$ Mpc, as a function of the magnetic field spectral index $n_B$. The bound is obtained from the effect on the CMB of the inflationary magnetic mode Eq.~\eqref{e:Clim}: it applies to magnetic fields generated during inflation. For $n_B>-3/2$, we have set $k_D\eta_0=3000$ (c.f. Ref.~\cite{Caprini:2009vk}).}
\end{figure}

A rough constraint on $B_\la$ as a function of $n_B$ can be derived by imposing that the amplitude of the CMB spectrum from the magnetic field at large scales must not overcome the observed value. Using the above equation~\eqref{zetaBlambda} into the CMB spectrum originated from the inflationary magnetic mode Eq.~\eqref{e:Clim}, and setting 
\be \label{e:Cltrue}
\ell^2 C_\ell^{\rm im} \leq \ell^2 C_\ell \simeq 7.9 \, \times \,10^{-10}\, , 
\ee
we derive the constraint shown in Fig.~\ref{fig:Blambda}. 

This figure shows that only in the nearly scale invariant case, where $n_B\sim -3$ it is possible to have large (nG) magnetic fields on the scale of $\lambda=1$ Mpc without generating too large a contribution from the inflationary magnetic mode in the CMB. This is due to the huge factor
\be
\frac{\la}{\eta_*} \simeq 1.4 \times 10^{23} \frac{T_*}{10^{16}{\rm GeV}} \frac{\la}{{\rm Mpc}}  \,,
\label{laetastar}
\ee
which enters in the amplitude of the inflationary mode when $\zeta_*$ is expressed in terms of $B_\la$ (c.f. Eq.~\eqref{zetaBlambda}). If $n_B<-3/2$, the constraint on $B_\la$ varies as $(\eta_*/\la)^{(n_B+3)/2}$. Therefore,  it becomes more stringent as the spectral index increases. If $n_B> -3/2$, the dependence on the spectral index is much weaker, since $B_\lambda$ varies as $(\eta_*/\la)^{3/4}/(k_D\la)^{(2n_B+3)/4} = (\eta_*/k_D)^{3/4}/(k_D\la)^{n_B/2}$ with $k_D^{-1}\gg \eta_*$. This change in the slope is clearly visible in Fig.~\ref{fig:Blambda}. 

The dependence on $B_\la$ and on $\eta_*$ is a general feature of the CMB spectrum generated by an inflationary magnetic field, and it does not depend on the details of the generation model that we have chosen in this analysis. 

A somewhat stronger constraint is obtained when taking into account that the CMB fluctuations from a magnetic field are non-Gaussian and lead to a bispectrum $B_\ell^{\rm im}$ which is parametrically of the order of
$\left(C_\ell^{\rm im}\right)^{3/2}$ such that 
$$
f_{\rm NL}\sim \ell^3B^{\rm im}_\ell/(\ell^2C_\ell)^2 \sim \left(\ell^2C_\ell^{\rm im}\right)^{3/2}/(\ell^2C_\ell)^2 \lsim 10\,. 
$$
The upper limit on $f_{\rm NL}$ is the Planck limit from~\cite{Ade:2013xxiv} on $f_{\rm NL}^{\rm local}$ .
Inserting (\ref{e:Cltrue}) for $\ell^2C_\ell$, we obtain
\be
\ell^2C_\ell^{\rm im}\ \lsim 4\times 10^{-3} \; \ell^2C_\ell \,.
\ee
Since $C_\ell^{\rm im}$ scales with $B_\la^4$ this reduces the limit shown in Fig.~\ref{fig:Blambda} only by about a factor 5.  On the other hand, this is a very rough estimate taking into account only the parametric dependence. The true value may well contain a pre-factor which differs considerably from $1$, see discussion of Ref.~\cite{barnaby} below. We therefore have plotted only the much safer limit $C_\ell^{\rm im} \le C_\ell$.

Strictly speaking, the CMB only puts a constraint on the parameter $C_{\rm em}(\gamma)=-C_{\Pi}(\gamma)$. In the context of the particular model given by Eqs.~\eqref{actionem} and~\eqref{f},  this parameter is known once the value of $\gamma$ is fixed \cite{maginf}. The CMB constraint can therefore be translated into a constraint, e.g. on the energy scale of inflation \cite{energyscale}. In this case, the amount of magnetic field generated is therefore completely determined by the choice of $\gamma$. It has been shown in previous analyses (and can be inferred from Eq. \eqref{ombprad}) that if $n_B$ is significantly larger than $-3$, $\omb$ is strongly suppressed on large scales and one cannot expect a large field amplitude $B_\la$ for $\la$ of the order of the Mpc \cite{yokoyama,subramanian}. However, the spirit of this paper is to put the inflationary magnetic mode on the same footing as the passive and the compensated modes, and to use it to set model independent constraints on the magnetic field amplitude today using the CMB. 

The constraint on $B_\lambda$ shown in Fig.~\ref{fig:Blambda} is model independent in the sense that it is mainly set by the dependence of Eq.~\eqref{zetaBlambda} on the factor $\lambda/\eta_*$. The numerical pre-factor in Eq.~\eqref{zetaBlambda} depends somewhat on the choice of the model of Eqs.~\eqref{actionem} and~\eqref{f}, but this dependence is negligible compared to the main feature of the constraint, i.e., its strong dependence on $n_B$. 

Basically, the difference from the passive and compensated modes is that, for the inflationary mode, the {\it integrated} magnetic energy density (up to the tiny scale $\eta_*$) is converted into a scale invariant $\zeta_*$, and influences the CMB constraint on the magnetic field amplitude at large scale $B_\lambda$. The passive and compensated modes, on the other hand, depend on the magnetic field spectrum: therefore only the large scales contribute to the CMB constraint. 

\begin{figure}[ht]
\centerline{\epsfig{figure=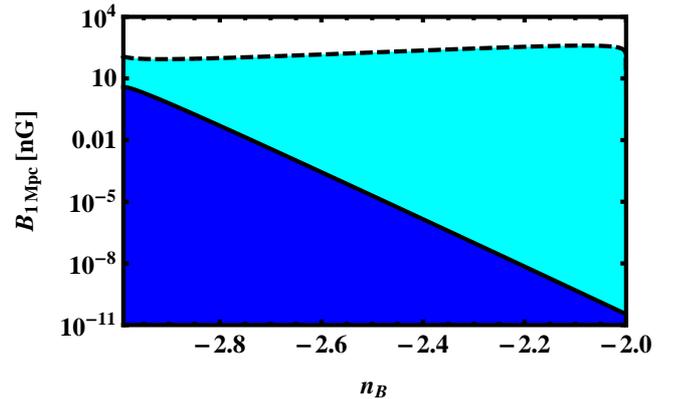,height=5.3cm}}
\caption{ \label{fig:zetapassive} Region below the dashed line: constraint from the passive mode from the CMB spectrum \eqref{Clpassivered}. Region below the solid line: the same as Fig.~\ref{fig:Blambda}. }
\end{figure}

In Fig.~\ref{fig:zetapassive} we compare the CMB constraint on $B_\lambda$ obtained from the passive and the inflationary mode, for $n_B<-2$ (for the passive CMB spectrum we use approximation \eqref{Clpassivered}). The constraint from the passive mode does not become more stringent for higher values of the spectral index. The CMB spectrum from the passive mode in terms of $B_\lambda $ contains in fact only a factor $(\lambda/\eta_0)^{(n_B+3)/2}$, instead of the huge factor of Eq.~\eqref{laetastar} \footnote{Note that we could have chosen to express the CMB constraint in terms of the integrated magnetic field amplitude $\vev{B^2}\simeq B_\lambda^2(k_D\lambda)^{n_B+3}$. We have chosen $B_\lambda$ since it is the quantity customarily used in CMB analyses. In terms of $\sqrt{\vev{B^2}}$, the result in Fig.~\ref{fig:Blambda} would simply change by a factor $(k_D \lambda)^{(n_B+3)/2}$.}. 
 
In a full MCMC study, one has to vary the parameters $\eta_*$, $B_{1{\rm Mpc}}$ and $n_B$. Note that the above equation~\eqref{zetaBlambda} can be used to express the initial conditions for the Boltzmann hierarchy given in Appendix \ref{app:initial} in terms of $B_{1{\rm Mpc}}$ and $n_B$. In the simple model discussed here, the additional slow roll parameter $\ep$ appearing in Eq.~\eqref{zetaBlambda} is actually not a free parameter but can be inferred from $B_{1{\rm Mpc}}$ and the amplitude of inflationary perturbations. However this may change in a more sophisticated model for the generation of magnetic fields during inflation.

As mentioned before, in addition to the power spectrum, the inflationary magnetic mode induces a distinct bispectrum since it is non-Gaussian~\cite{barnaby,Caprini:2009vk}. As for the CMB spectrum, there is a non-Gaussian contribution in the CMB temperature anisotropy arising from the inflationary magnetic mode (the one calculated in \cite{barnaby}) and one arising from the passive and compensated modes, which corresponds to the one evaluated in e.g. \cite{Caprini:2009vk}. If the generation of the magnetic field arises during inflation, the non-Gaussian contribution from the inflationary magnetic mode is the dominant one. Ref.~\cite{barnaby} estimates that a scale invariant magnetic field spectrum generates a bi-spectrum with $f_{\rm NL}^{\rm equiv.~local} \simeq 1280 \,\mathcal{P} \, N_{\rm CMB}^3(N_{\rm tot}-N_{\rm CMB})$, where $\mathcal{P}=H_*^2/(\pi m_P^2\ep) \simeq 2.1\times 10^{-9}$ is the amplitude of the fluctuations of the inflationary spectrum, $N_{\rm CMB}$ is the number of e-folds before the end of inflation when the observable scales leave the horizon, while $N_{\rm tot}$ is the total number of e-folds of inflation\footnote{Since the calculations of $\zeta$ in Refs.~\cite{barnaby} and \cite{maginf} differ significantly, we show in Appendix~\ref{app:curvature} that they are nevertheless equivalent.}. Because of the presence of $N_{\rm tot}$, in the perfectly scale invariant case the amount of non-Gaussianity cannot be determined precisely. On the other hand, for a magnetic field spectrum only close to scale invariance, the factor $(N_{\rm tot}-N_{\rm CMB})$ is absent, since it comes directly from the logarithmic divergence of the magnetic energy density power spectrum when $n_B=-3$. If the magnetic spectral index is close to scale invariance, we can assume for an order of magnitude estimate that the calculation of \cite{barnaby} applies also to this case, and that $f_{\rm NL}^{\rm equiv.~local}\sim 1280 \, \mathcal{P} \, N_{\rm CMB}^3$: one obtains then $f_{\rm NL}^{\rm equiv.~local} \simeq 0.4$ to $0.7$, well below the current Planck limit of roughly $f_{\rm NL} \le 10$~\cite{Ade:2013xxiv}. There is therefore no indication that the model analyzed in this paper is excluded by the new bounds on non-Gaussianity from Planck. Constraints on the magnetic field amplitude $B_{1{\rm Mpc}}$ as a function of the spectral index can be placed imposing the Planck upper bound on $f_{\rm NL}$ on the non-Gaussianity due to the inflationary magnetic mode. As for the constraints from the power spectrum, we expect these would be stronger than the constraints derived in \cite{Caprini:2009vk} from the compensated mode.

\section{Conclusions}
\label{sec:con}

In this paper we have computed the CMB anisotropy spectrum from magnetic fields generated during inflation. We have paid special attention to a new mode which we call the inflationary magnetic mode. It is due to contributions to the comoving curvature perturbation $\zeta$ which come from the perturbations of the geometry by the magnetic field during inflation. 

This mode is always scale invariant, and dominates the CMB signal with respect to the passive and compensated magnetic modes, which are sourced by the magnetic field after inflation. We have evaluated analytically the CMB spectra from the inflationary magnetic mode and inferred an analytical constraint on the magnetic field amplitude today, as a function of the magnetic field spectral index. The constraint is much stronger than what is usually found with CMB analyses for spectral indexes far from scale invariance: through the inflationary magnetic mode, even a magnetic field with spectrum far from scale invariance can leave a detectable imprint in the CMB. The new mode should therefore be accounted for when constraining primordial magnetism with the CMB. This implies however that the magnetic field generation time must be inserted as a new parameter in CMB analyses, and the constraints on the magnetic field amplitude and spectral index should be diversified depending on the generation mechanism of the primordial magnetic field. 

In this analysis we started from a given model of inflationary magnetogenesis, where the standard electromagnetic action is modified by inserting a specific coupling with the inflaton in the kinetic term, breaking conformal invariance. However, the constraint on the magnetic field amplitude that we finally derive from the CMB temperature spectrum does not depend strongly on the choice of the magnetogenesis model, apart from the numerical pre-factor. The strong dependence on the magnetic spectral index, which is the novelty of this CMB constraint with respect to those obtained using the passive and compensated modes, is a general feature of any magnetic mode in the curvature perturbation generated during inflation. 

Magnetic fields also generate vector and tensor perturbations during inflation: a homogeneous (inflationary) magnetic vector mode subsequently decays and does not leave any signature in the CMB, however, a tensor mode might be present and should also be taken into account in a full MCMC study. 

Finally, we want to point out that the inflationary magnetic mode obtained in Ref.~\cite{maginf} is actually more general than its derivation. The vacuum fluctuations of an arbitrary light field which is not conformally coupled, e.g. a minimally coupled light scalar field, will be amplified during inflation. Even if the field decays after inflation e.g. into standard model particles, its contribution $\zeta_*$ to the curvature will remain by continuity and might  result in observable temperature anisotropies which, generically will be non-Gaussian. Therefore, limits on primordial non-Gaussianity also provide a limit on the number of light fields which are non-conformally coupled during inflation. This is an interesting new window into very high energy physics from cosmological observations. 

\acknowledgments{ 
It is a pleasure to thank Fabio Finelli, Lukas Hollenstein, Daniela Paoletti, Levon Pogosian and Federico Urban for very useful discussions, and Richard Shaw for his help with the magnetic version of CAMB. CB is supported by the Herchel Smith Postdoctoral fund and
by King's College Cambridge. 
RD is supported by the Swiss National Science Foundation and the (US) National Science Foundation under Grant No. NSF PHY11-25915. }

\appendix

\section{Initial conditions for the new mode}
\label{app:initial}
We can derive initial conditions for the metric and matter perturbations variables by combining Einstein's and the conservation equations. We place ourselves in the radiation era, well after neutrino decoupling so that the anisotropic stress of the magnetic field has been completely compensated by the neutrino anisotropic stress at lowest order: $\pi_\nu=-3\omp/R_\nu$. We give the initial conditions for the synchronous gauge variables used in CAMB and for the gauge invariant metric potentials $\Phi$ and $\Psi$. The passive and compensated modes are as in~\cite{shaw, finelli}. Here we write only the initial conditions for the inflationary magnetic mode $\zeta_*$ (an expression for $\zeta_*$ in terms of the magnetic field parameters is given in Eq.~\eqref{zetaBlambda}):
\bea
h(\eta)&=&\frac{1}{2}(k\eta)^2\zeta_*\\
\eta(\eta)&=&\left( 1-\frac{5+4R_\nu}{12(15+4R_\nu)}\right)(k\eta)^2\zeta_*\\
\delta_c(\eta)&=&-\frac{1}{4}(k\eta)^2\zeta_*\\
v_c&=&0\\
\delta_\nu&=&-\frac{1}{3}(k\eta)^2\zeta_*\\
v_\nu&=&-\frac{23+4R_\nu}{36(15+4R_\nu)}(k\eta)^3\zeta_*\\
\pi_\nu&=&\frac{4}{15+4R_\nu}(k\eta)^2\zeta_*\\
F_{\nu3}&=&\frac{4}{3(15+4R_\nu)}(k\eta)^3\zeta_*\\
\delta_b(\eta)&=&-\frac{1}{4}(k\eta)^2\zeta_*\\
v_b&=&-\frac{1}{36}(k\eta)^3\zeta_*\\
\delta_\gamma(\eta)&=&-\frac{1}{3}(k\eta)^2\zeta_*\\
v_\gamma&=&-\frac{1}{36}(k\eta)^3\zeta_*\\
\Psi&=&\frac{2(5+2R_\nu)}{15+4R_\nu}\zeta_*\\
\Phi&=&\frac{10}{15+4R_\nu}\zeta_*
\eea

\section{The comoving curvature perturbation}
\label{app:curvature}

In this appendix we compare the approaches of Ref.~\cite{maginf} and Ref.~\cite{barnaby} to calculate the comoving curvature perturbation $\zeta$. Ref.~\cite{maginf} derives and solves directly the evolution equation for $\zeta$, while Ref.~\cite{barnaby} solves the Klein-Gordon equation for the inflaton field perturbation $\delta\varphi$ in the flat gauge. The two procedures must be exactly equivalent. However, this is not immediately apparent, since the source term of the equation for the inflaton field perturbation (c.f. Eq. (43) of \cite{barnaby}) appears at first sight unrelated to the source of the equation for $\zeta$ (c.f. Eq.~(17) of \cite{maginf}), provided by the electromagnetic energy density $\rho_\eb$ and anisotropic stress $\Pi={T^i}_i/2-3/2 \hat{k}^i\hat{k}^j{T^j}_i$ (${T^\mu}_\nu$ is the energy momentum tensor of the electromagnetic field):
\bea
& \ddot{\delta\varphi}+2\HH \dot{\delta\varphi}+k^2\delta\varphi=a^2 f \frac{df}{d\bar\varphi} (E^2-B^2) \label{eqphi}\\
& \ddot{\zeta}+2\HH \dot{\zeta}+k^2\zeta=-\frac{2}{3}\frac{\HH^2 a^2}{\dot{\bar\varphi}^2}\left[ 6\rho_\eb+\frac{\dot{\rho}_\eb}{\HH}+\frac{\dot\Pi}{\HH}\right] 
\label{eqzetacom}
\eea
where $\bar\varphi$ is the background inflaton field. In order to demonstrate that these two equations are indeed equivalent, one has to derive the relation between the comoving curvature perturbation $\zeta$ and the scalar field perturbation in flat gauge $\de\varphi$. Extending the definitions given in~\cite{Malik:2008im} to incorporate the electromagnetic field, one finds 
\be
\zeta=\frac{\HH}{\dot{\bar\varphi}} \,\de\varphi  +\frac{\HH a^2}{\dot{\bar\varphi}^2} \, q_\eb
\label{zetadeltaphi}
\ee
where $q_\eb=-{\rm i}\hat{k}^i\,{T^0}_i / k$ and ${T^0}_i$ is the Poynting vector component of the electromagnetic energy-momentum tensor. In order to demonstrate the equivalence of Eqs.~\eqref{eqphi} and \eqref{eqzetacom}, it is enough to derive the above expression twice and make use of the electromagnetic energy and momentum conservation equations, that read:
\bea
& \dot\rho_\eb+4\HH\rho_\eb+k^2 q_\eb= \dot{\bar\varphi} \, f \frac{df}{d\bar\varphi} (B^2-E^2) \label{emencons}\\
& \dot q_\eb+4\HH q_\eb-\frac{\rho_\eb}{3}+\frac{2}{3}\Pi =0
\eea
The source term of the Klein Gordon equation \eqref{eqphi} is related to the energy conservation of the system composed by the scalar field plus the electromagnetic field: from the definition of the total energy momentum tensor in terms of the total Lagrangian $\mathcal{L}_\eb+\mathcal{L}_\varphi$ it is easy to show that
\be
\nabla^\mu T_{\mu\nu}=\frac{\partial \mathcal{L}_\eb}{\partial\varphi}\nabla_\nu\varphi=\nabla_\mu\left( \frac{\partial \mathcal{L}_\varphi}{\partial( \nabla_\mu\varphi)} \right) \nabla_\nu\varphi
\ee
where $T_{\mu\nu}$ denotes as usual the electromagnetic energy momentum tensor and the second equality comes from the Klein Gordon equation. 

The two approaches of Ref.~\cite{maginf} and Ref.~\cite{barnaby} are therefore equivalent. Solving Eq. \eqref{eqzetacom} seems to us more convenient, since one has direct access to the metric perturbation, which is the relevant quantity to evaluate directly the CMB anisotropies. In the standard inflationary case, once the field perturbation in the flat gauge $\de\varphi$ is determined, so is the comoving curvature perturbation $\zeta$; when an electromagnetic field is present, an extra term (the Poyting vector) relates the two quantities, c.f. \eqref{zetadeltaphi}. Ref.~\cite{barnaby} shows a posteriori that this term is anyway negligible for $\zeta$ with respect to the contribution from $\de\varphi$. Note that there is in principle no need to involve in the calculation the curvature perturbation in the uniform density gauge.


\begin{thebibliography}{99}

\bibitem{Puy:1998sv}
  D.~Puy and P.~Peter,
  Int.\ J.\ Mod.\ Phys.\ D {\bf 07} (1998) 489
  [astro-ph/9802329].

\bibitem{Jedamzik:1999bm}
  K.~Jedamzik, V.~Katalinic and A.~V.~Olinto,
  Phys.\ Rev.\ Lett.\  {\bf 85} (2000) 700
  [astro-ph/9911100].

 \bibitem{Durrer:1999bk}
  R.~Durrer, P.~G.~Ferreira and T.~Kahniashvili,
  Phys.\ Rev.\ D {\bf 61} (2000) 043001
  [arXiv:astro-ph/9911040].
 
\bibitem{Caprini:2003vc}
 C. Caprini, R. Durrer, Ruth and T.~Kahniashvili, 
  Phys.Rev.{\bf D69} (2004) 063006.
  [arXiv:astro-ph/0304556].
   
 \bibitem{Paoletti:2012bb}
  D.~Paoletti and F.~Finelli,
  arXiv:1208.2625 [astro-ph.CO].
  
  \bibitem{Paoletti:2010rx}
  D.~Paoletti and F.~Finelli,
  Phys.\ Rev.\ D {\bf 83} (2011) 123533
  [arXiv:1005.0148 [astro-ph.CO]].
  
  \bibitem{finelli}
  D.~Paoletti, F.~Finelli, F.~Paci,
  Mon.\ Not.\ Roy.\ Astron.\ Soc.\  {396 } (2009)  523-534.
  [arXiv:0811.0230 [astro-ph]];\\ 
  F. Finelli, F. Paci and D. Paoletti, Phys. Rev. D78, 023510 (2008).
  
\bibitem{Shaw:2010ea}
  J.~R.~Shaw and A.~Lewis,
  Phys.\ Rev.\ D {\bf 86} (2012) 043510
  [arXiv:1006.4242 [astro-ph.CO]].

\bibitem{shaw} J.~R.~Shaw and A.~Lewis, Phys. Rev. {D81},  043517 (2010).  

  \bibitem{Lewis:2004ef}
  A.~Lewis,
  Phys.\ Rev.\ D {\bf 70} (2004) 043011
  [astro-ph/0406096].
  
  \bibitem{Kunze:2010ys}
  K.~E.~Kunze,
  Phys.\ Rev.\ D {\bf 83} (2011) 023006
  [arXiv:1007.3163 [astro-ph.CO]].
  
  \bibitem{Giovannini:2009zq}
  M.~Giovannini,
  Phys.\ Rev.\ D {\bf 79} (2009) 121302
  [arXiv:0902.4353 [astro-ph.CO]].
  
  \bibitem{Giovannini:2007qn}
  M.~Giovannini and K.~E.~Kunze,
  Phys.\ Rev.\ D {\bf 77} (2008) 063003
  [arXiv:0712.3483 [astro-ph]].
    
  \bibitem{Yamazaki:2011eu}
  D.~G.~Yamazaki, K.~Ichiki, T.~Kajino and G.~J.~Mathew,
  Adv.\ Astron.\  {\bf 2010} (2010) 586590
  [arXiv:1112.4922 [astro-ph.CO]].
  
  \bibitem{Yamazaki:2012pg}
  D.~G.~Yamazaki, T.~Kajino, G.~J.~Mathew and K.~Ichiki,
  Phys.\ Rept.\  {\bf 517} (2012) 141
  [arXiv:1204.3669 [astro-ph.CO]].
  
\bibitem{Durrer:2006pc} Durrer, Ruth,
 New Astron. Rev., {\bf 51}, (2007) 275.
 [arXiv:astro-ph/0609216].

  \bibitem{Seshadri:2009sy}
  T.~R.~Seshadri and K.~Subramanian,
  Phys.\ Rev.\ Lett.\  {\bf 103} (2009) 081303
  [arXiv:0902.4066 [astro-ph.CO]].

 \bibitem{Caprini:2009vk}
  C.~Caprini, F.~Finelli, D.~Paoletti and A.~Riotto,
  JCAP {\bf 0906} (2009) 021
  [arXiv:0903.1420 [astro-ph.CO]].

    \bibitem{Shiraishi:2012rm}
  M.~Shiraishi, D.~Nitta, S.~Yokoyama and K.~Ichiki,
  JCAP {\bf 1203} (2012) 041
  [arXiv:1201.0376 [astro-ph.CO]].
  
  \bibitem{Shiraishi:2010yk}
  M.~Shiraishi, D.~Nitta, S.~Yokoyama, K.~Ichiki and K.~Takahashi,
  Phys.\ Rev.\ D {\bf 82} (2010) 121302
   [Erratum-ibid.\ D {\bf 83} (2011) 029901]
  [arXiv:1009.3632 [astro-ph.CO]].
  
  
  \bibitem{Cai:2010uw}
  R.~-G.~Cai, B.~Hu and H.~-B.~Zhang,
  JCAP {\bf 1008} (2010) 025
  [arXiv:1006.2985 [astro-ph.CO]].
    
    \bibitem{Kosowsky:2004zh}
  A.~Kosowsky, T.~Kahniashvili, G.~Lavrelashvili and B.~Ratra,
  Phys.\ Rev.\ D {\bf 71} (2005) 043006
  [astro-ph/0409767].
    
  \bibitem{Yadav:2012uz}
  A.~Yadav, L.~Pogosian and T.~Vachaspati,
  Phys.\ Rev.\ D {\bf 86} (2012) 123009
  [arXiv:1207.3356 [astro-ph.CO]].
  
  
 \bibitem{Ade:2013zuv}
  P.~A.~R.~Ade {\it et al.}  [The Planck Collaboration, 2013 results XVI],
  arXiv:1303.5076 [astro-ph.CO].
  
  
  \bibitem{Jedamzik:1996wp}
  K.~Jedamzik, V.~Katalinic and A.~V.~Olinto,
  Phys.\ Rev.\ D {\bf 57} (1998) 3264
  [astro-ph/9606080].
  
  \bibitem{Subramanian:1997gi}
  K.~Subramanian and J.~D.~Barrow,
  Phys.\ Rev.\ D {\bf 58} (1998) 083502
  [astro-ph/9712083].
  
  \bibitem{ruthreview} 
  R.~Durrer and A.~Neronov,
  arXiv:1303.7121 [astro-ph.CO].

\bibitem{Durrer:2003ja}
  R.~Durrer and C.~Caprini,
  JCAP {\bf 0311} (2003) 010
  [astro-ph/0305059].
  
  \bibitem{demozzi}
 V.~Demozzi, V.~Mukhanov, H.~Rubinstein,
  JCAP {0908 } (2009)  025.
  [arXiv:0907.1030 [astro-ph.CO]].
  
  \bibitem{Himmetoglu:2009qi}
  B.~Himmetoglu, C.~R.~Contaldi and M.~Peloso,
  Phys.\ Rev.\ D {\bf 80} (2009) 123530
  [arXiv:0909.3524 [astro-ph.CO]].
  
\bibitem{Giovannini:2008yz}
  M.~Giovannini and K.~E.~Kunze,
  Phys.\ Rev.\ D {\bf 77} (2008) 123001
  [arXiv:0802.1053 [astro-ph]].

\bibitem{magcaus} C. Bonvin and C. Caprini, JCAP {1005}, 022 (2010) 
 [arXiv:1004.1405].
 
\bibitem{maginf} C. Bonvin, C. Caprini and R. Durrer,   Phys. Rev. {D86}, 023519 (2012) 
 [arXiv:1112.3901].
 
\bibitem{barnaby}
  N.~Barnaby, R.~Namba and M.~Peloso,
  Phys.\ Rev.\ D {\bf 85} (2012) 123523
  [arXiv:1202.1469 [astro-ph.CO]].
  
   \bibitem{seery} D. Seery, JCAP 0908:018 (2009) [arXiv:0810.1617 [astro-ph]].
  
  \bibitem{mukhanov} V.F. Mukhanov, H.A. Feldman and R.H. Brandenberger, Phys. Rep. {215}, 203 (1992).
    
\bibitem{turner} M.S. Turner and L.M. Widrow, Phys. Rev. {D37}, 2743 (1988).

\bibitem{ratra} B. Ratra, Astrophys. J. Lett. {391}, L1 (1992).
  
   \bibitem{Giovannini:2000dj}
  M.~Giovannini and M.~E.~Shaposhnikov,
  Phys.\ Rev.\ D {\bf 62} (2000) 103512
  [hep-ph/0004269].
  
 \bibitem{Dimopoulos:2001wx}
  K.~Dimopoulos, T.~Prokopec, O.~Tornkvist and A.~C.~Davis,
  Phys.\ Rev.\ D {\bf 65} (2002) 063505
  [astro-ph/0108093].
  
\bibitem{bamba1}  K.~Bamba and J.~Yokoyama,
  Phys.\ Rev.\  D {69} (2004) 043507
  [arXiv:astro-ph/0310824].
  
  \bibitem{Anber:2006xt}
  M.~M.~Anber and L.~Sorbo,
  JCAP {\bf 0610} (2006) 018
  [astro-ph/0606534].
  
\bibitem{bamba2}  K.~Bamba, N.~Ohta, S.~Tsujikawa,
  Phys.\ Rev.\  {D78 } (2008)  043524.
  [arXiv:0805.3862 [astro-ph]].
    
   \bibitem{yokoyama} J.~Martin, J.~'i.~Yokoyama,
  JCAP {0801 } (2008)  025
  [arXiv:0711.4307 [astro-ph]].
   
     \bibitem{subramanian} K. Subramanian, Astron.Nachr. {331}, 110 (2010) [arXiv:0911.4771]. 

   
  \bibitem{Durrer:2010mq} 
R. Durrer, L. Hollenstein and R.K. Jain,
     JCAP, {1103}, 037 (2011) [arXiv:1005.5322].

   \bibitem{Ferreira:2013sqa}
  R.~J.~Z.~Ferreira, R.~K.~Jain and M.~S.~Sloth,
  arXiv:1305.7151 [astro-ph.CO].
  
  \bibitem{Barnaby:2011vw}
  N.~Barnaby, R.~Namba and M.~Peloso,
  JCAP {\bf 1104} (2011) 009
  [arXiv:1102.4333 [astro-ph.CO]].
  
 

  
  \bibitem{AE} J.~Ahonen \& K.~Enqvist, {\it Phys. Lett.} {B382}, 40 (1996);\\
    G.~Baym \& H.~Heiselberg, {\it Phys. Rev.} {D56}, 5254 (1997).
    



\bibitem{muk_deruelle} 	N. Deruelle and V.F. Mukhanov, Phys. Rev. {D52}, 5549 (1995); \\
	N. Deruelle, D. Langlois and J.-P. Uzan, Phys. Rev. { D56}, 7608  (1997). 

\bibitem{Kojima:2009gw}
K.~Kojima, T.~Kajino, G.~J.~Mathews,
JCAP {\bf 1002} (2010) 018,
arXiv:0910.1976 [astro-ph.CO] 

\bibitem{const} J. Adamek, R. Durrer, E. Fenu and M. Vonlanthen, JCAP06(2011)017 [arXiv:1102.5235].

\bibitem{mybook}R. Durrer, {\em The Cosmic Microwave Background}, Cambridge 
         University Press (Cambridge, 2008).


\bibitem{xx}
C.~J.~Hogan, Phys.\ Rev.\ Lett.\ {\bf 51}, (1983) 1488;
%
J.~M.~Quashnock, A. Loeb and D. N. Spergel, Astrophys.\ J.\ {\bf 344} (1989) L49;
%
B.~Cheng and A.V.~Olinto, Phys.\ Rev.\  D {\bf 50} (1994) 2412;
%
G.~Baym, D. B\"odeker and L. McLerran, Phys.\ Rev.\  D {\bf 53} (1996) 662;
%
G.~Sigl and A. V. Olinto., Phys.\ Rev.\  D {\bf 55} (1997) 4582;
%
  D.~Boyanovsky and H.~J.~de Vega,
  AIP Conf.\ Proc.\  {\bf 784} (2005) 434
  [arXiv:astro-ph/0502212];
%
T. Vachaspati, Phys.\ Lett.\  B {\bf 265} (1991) 258;
%
  K.~Enqvist and P.~Olesen,
  Phys.\ Lett.\  B {\bf 319} (1993) 178
  [arXiv:hep-ph/9308270];
%
  M.~Hindmarsh and A.~Everett,
  Phys.\ Rev.\  D {\bf 58} (1998) 103505
  [arXiv:astro-ph/9708004];
%
D.~Grasso and A.~Riotto,
  Phys.\ Lett.\  B {\bf 418} (1998) 258
  [arXiv:hep-ph/9707265];
%
J.~M.~Cornwall, Phys.\ Rev.\  D {\bf 56} (1997) 6146;
%
T.~Vachaspati, Phys.\ Rev.\ Lett.\  {\bf 87} (2001) 251302;
%
  M.~Joyce and M.~E.~Shaposhnikov,
  Phys.\ Rev.\ Lett.\  {\bf 79} (1997) 1193
  [arXiv:astro-ph/9703005];
%
  G.~B.~Field and S.~M.~Carroll,
  Phys.\ Rev.\  D {\bf 62} (2000) 103008
  [arXiv:astro-ph/9811206];
%
  L.~Campanelli and M.~Giannotti,
  Phys.\ Rev.\  D {\bf 72} (2005) 123001
  [arXiv:astro-ph/0508653].
  
\bibitem{energyscale} V.~Demozzi and C.~Ringeval,
  JCAP {\bf 1205} (2012) 009
  [arXiv:1202.3022 [astro-ph.CO]]; C.~Ringeval, T.~Suyama and J.~'i.~Yokoyama,
  arXiv:1302.6013 [astro-ph.CO].


  
 \bibitem{Ade:2013xxiv}
  P.~A.~R.~Ade {\it et al.}  [The Planck Collaboration, 2013 results XXIV],
  arXiv:1303.5084 [astro-ph.CO].
  
\bibitem{Malik:2008im}
  K.~A.~Malik and D.~Wands,
  Phys.\ Rept.\  {\bf 475} (2009) 1
  [arXiv:0809.4944 [astro-ph]].


\end{thebibliography}
\end{document}